\documentclass[table,sigconf]{acmart}
\usepackage{hyperref}
\usepackage{float}
\usepackage{multirow}
\usepackage{listings}
\usepackage{fp}
\usepackage{xparse}

\newcount\ppnum
\newcommand\numc[1]{%
        \ppnum=#1\relax
        \ifnum\ppnum<0
                $-$%
                \ppnum=-\ppnumx
        \fi
        \let\pptemp\empty
        \loop\ifnum\ppnum>9999
                \count255=\ppnum
                \divide\ppnum by1000
                \count255=\numexpr \count255 - 1000*\ppnum \relax
                \edef\pptemp{,\ifnum\count255<100 0\ifnum\count255<10 0\fi\fi
                             \the\count255 \pptemp}%
        \repeat
        \the\ppnum
        \pptemp
}

\newcommand\new[1]{
  \IfNoValueF{#1}{{\color{black} #1}}
}

\newenvironment{newtext}{\par\color{black}}{\par}

\newcommand{\subhead}[1]{\vspace{0pt}\noindent{\textbf{#1.}}}

\newcommand{\TechniqueName}{\textit{custom encryption}}
\newcommand{\ToolName}{\textit{ThirdEye}}
\newcommand\printpercent[2]{\numc{#1} (\FPeval\result{round(#1*100/#2,2)}\result\%)}
\newcommand\printpercentoutof[2]{\numc{#1}/\numc{#2} (\FPeval\result{round(#1*100/#2,2)}\result\%)}

\copyrightyear{2022}
\acmYear{2022}
\setcopyright{acmcopyright}\acmConference[CCS '22]{Proceedings of the 2022 ACM
SIGSAC Conference on Computer and Communications Security}{November 7--11,
2022}{Los Angeles, CA, USA}
\acmBooktitle{Proceedings of the 2022 ACM SIGSAC Conference on Computer and
Communications Security (CCS '22), November 7--11, 2022, Los Angeles, CA, USA}
\acmPrice{15.00}
\acmDOI{10.1145/3548606.3560665}
\acmISBN{978-1-4503-9450-5/22/11}

\begin{document}

\title[\emph{Hidden in Plain Sight}: Exploring Encrypted Channels in Android apps]{\emph{Hidden in Plain Sight}: \\Exploring Encrypted Channels in Android apps}\thanks{This is an extended version of an ACM CCS 2022 paper with the same title.}

\author{Sajjad Pourali, Nayanamana Samarasinghe, Mohammad Mannan}
\affiliation{
 \institution{Concordia University}
 \country{Canada}}
\email{{s\_poural,n\_samara, mmannan}@ciise.concordia.ca}

\renewcommand{\shortauthors}{Pourali et al.}

\begin{abstract}
As privacy features in Android operating system improve, privacy-invasive apps may gradually shift their focus to non-standard and covert channels for leaking private user/device information. Such leaks also remain largely undetected by state-of-the-art privacy analysis tools, which are very effective in uncovering privacy exposures via regular HTTP and HTTPS channels. In this study, we design and implement, \textit{\ToolName}, to significantly extend the visibility of current privacy analysis tools, in terms of the exposures that happen across various non-standard and covert channels, i.e., via any protocol over TCP/UDP (beyond HTTP/S), and using multi-layer custom encryption over HTTP/S and non-HTTP protocols. Besides network exposures, we also consider covert channels via storage media that also leverage custom encryption layers. Using \ToolName, we analyzed \numc{12598} top-apps in various categories from Androidrank, and found that \printpercentoutof{2887}{12598} apps used custom encryption/decryption for network transmission and storing content in shared device storage, and \printpercentoutof{2465}{2887} of those apps sent device information (e.g., advertising ID, list of installed apps) over the network that can fingerprint users. Besides, \numc{299} apps transmitted insecure encrypted content over HTTP/non-HTTP protocols; \numc{22} apps that used authentication tokens over HTTPS, happen to expose them over insecure (albeit custom encrypted) HTTP/non-HTTP channels. We found non-standard and covert channels with multiple levels of obfuscation (e.g., encrypted data over HTTPS, encryption at nested levels), and the use of vulnerable keys and cryptographic algorithms. Our findings can provide valuable insights into the evolving field of non-standard and covert channels, and help spur new countermeasures against such privacy leakage and security issues.
\end{abstract}



\maketitle

\section{Introduction}

Many Android apps collect a lot of privacy-sensitive information from users and share it with multiple parties, e.g., app servers, ad/tracking companies. Such data collection and sharing, leading to privacy breaches, has been extensively studied~\cite{50_ways,yang2013appintent,naseri2019accessileaks,pham2019hidemyapp}. However, these studies mostly deal with information exposure via insecure channels (e.g., incorrect deployment of HTTPS, or using HTTP), or via the standard HTTPS channel. On the other hand, some apps use additional side/covert channels, standard and non-standard protocols, with/without additional encryption layers (\emph{custom encryption}), for data transmission. The privacy profile of these apps (some of which are very popular) remains largely unscrutinized, even though some prominent examples exist (e.g., deceptive location-tracking by the ad company InMobi, fined by the US FTC~\cite{inmobi}).
Challenges of studying these channels include dealing with non-standard protocols (e.g., custom implementations over TCP/UDP), and detecting and decrypting
\mbox{custom encryption layers.}

Several studies~\cite{obfus_res,50_ways,spreitzer2018scandroid,block2017autonomic} have focused on the design, detection, and prevention of side and covert channels. Continella et al.~\cite{obfus_res} designed a framework to detect privacy leaks that is resilient to various obfuscation techniques (e.g., encoding, formatting). Reardon et al.~\cite{50_ways} looked into network traffic collected from apps/libraries to identify side and covert channels used to send sensitive information. Spreitzer et al.~\cite{spreitzer2018scandroid} developed an automated framework to detect side channel leaks (e.g., transmitted/received bytes, free/used space, CPU time) from Android APIs. Limitations of these studies include: lack of (or insufficient) support for non-HTTP protocols, custom encryption layers (beyond HTTPS),
and modern anti-reverse engineering evasion techniques; handling only a few fixed privacy-sensitive items (e.g., Android ID, contacts, IMEI, location, and phone number) sent over custom encrypted channels; shallow interaction with the apps (e.g., lack of sign-in/up support); and dependence on old/obsolete versions of Android.

By addressing the above limitations of state-of-the-art privacy analysis tools, we design and implement \textit{\ToolName}\footnote{In addition to permission checks and network flow monitoring, we add a \emph{third} perspective to observe app behaviors. In many Asian legends, the \emph{third eye} is meant to provide ``perception beyond ordinary sight'' -- see \url{https://en.wikipedia.org/wiki/Third_eye}.} that can effectively and automatically detect privacy exposures in non-standard channels over HTTP/HTTPS and non-HTTP protocols, where apps use one or more layers of \TechniqueName. We also consider custom encryption use and covert channels over storage media. \ToolName\ is designed for efficient and large-scale automated analysis, with real Android devices.
\new{
With \ToolName, we target the following goals in regards to the use of non-standard custom encryption channels: (i) to effectively and efficiently detect privacy leaks that occur through these channels; (ii) to identify security weaknesses introduced by these channels; (iii) to perform a measurement study of the prevalence of privacy leakage and security weaknesses in commonly used Android apps, due to these channels.
}

\ToolName\  is powered by four main components: the \emph{device manager} orchestrates app installs/launches/uninstalls on real Android devices, while maintaining the connection with a test desktop; the \emph{UI interactor} systematically traverses menus and UI items for comprehensive functionality coverage of an app; the \emph{operations logger} performs network/cryptographic/file API instrumentations using \textit{Frida} API hooking for capturing all inputs/outputs from these APIs; the \emph{data flow inspector} detects privacy and security issues in the collected network traffic/files. Besides privacy breaches, we also identify several security weaknesses in these non-standard channels, including: the use of fixed keys and weak/broken algorithms for encryption/decryption in files and network communication.

We implement \textit{\ToolName} on Android 12, which significantly extends privacy and security features compared to older versions; note that several past tools (designed for standard protocols primarily HTTP/S) are becoming much less effective or even incompatible on the newer versions of Android. Our UI interactor is more comprehensive and systematic than Android Monkey; we explore all UI elements based on their parameters and avoid visiting duplicate elements and pop-up advertisements/in-app purchases. The ability to handle automated sign-up and sign-in support (where possible) helps us cover app functionalities beyond the login prompt (where most tools stop). For improved automation on real Android devices, we provide comprehensive recovery from app crashing/freezing, and phone states that impede effective analysis (e.g., apps that can change WiFi settings). We address  common anti-evasion techniques (e.g., bypass root/package installer/mock location detection) to increase our app coverage. However, our implementation is currently unable to decode complex obfuscation, and protocols such as QUIC, DTLS; we also do not support app code in Android NDK.

Our implementation requires significant engineering efforts (approx.\ 5.5K and 1.5K LOC of Python and JavaScript code) to realize our design goals. We also leverage several existing tools including \textit{Frida}, \textit{Androguard}~\cite{androguard}, \textit{mitmproxy}~\cite{mitmproxy}, \textit{tcpdump}~\cite{tcpdump}, \textit{AndroidViewClient}~\cite{androidviewclient}, \textit{Python Translators Library}~\cite{translators} (for Google translation), \textit{adb}~\cite{adb} and Android internal commands. We mainly use Frida~\cite{Frida} to collect cryptographic parameters, trace shared storage and app-generated network traffic, and evade anti-reverse engineering mitigations. Additionally, we integrate Frida and Androguard to create a rule-based API logger that allows us to detect and collect non-SDK encryption/decryption APIs parameters. We use mitmproxy and tcpdump to capture HTTP/S and non-HTTP/S traffic, respectively. Our UI interactor is built on top of AndroidViewClient to traverse different objects, including buttons and inputs. We use the Google Translate API to enable support of non-English apps.

\vspace{2pt}
\noindent Our contributions and notable findings include:

\noindent(\textbf{1}) We design \textit{\ToolName} to find privacy and security exposures from various non-standard and covert channels. Unlike past work, \textit{\ToolName} can uncover privacy exposures and security issues
in both HTTP/HTTPS and non-HTTP protocols (i.e., protocols over TCP/UDP), and shared storage (on-device).

\noindent(\textbf{2}) Our implementation of \ToolName\ enables efficient, large-scale automated analysis of thousands of apps on real Android devices. We used  two Android devices (Pixel 4 and Pixel 6) running factory images with Android 12, to evaluate \numc{15522} top-apps from various categories in Androidrank~\cite{androidrank}; \numc{12598} (out of \numc{15522}, 81.2\%) apps were successfully analyzed (others failed for various download/compatibility issues). \ToolName\ successfully uncovered numerous novel privacy leakages and security issues.

\noindent(\textbf{3}) We identify \printpercentoutof{2887}{12598} apps use custom encryption/decryption for network transmission and storing content in the shared device storage;  \printpercentoutof{2383}{2887} of them transmit the on-device information that is commonly used for user tracking (e.g., advertising ID, router SSID, device email, list of the installed apps).
More importantly, for at least
one on-device info item, \printpercentoutof{2156}{2383} of the apps send it only under custom encryption to at least
one host, and \printpercentoutof{1719}{2383} apps send it only under custom encryption. %
All these serious privacy leakages would be missed by existing analysis tools.

\noindent(\textbf{4}) Besides privacy leakage, we also identify that the use of custom encryption channels can seriously undermine data security, e.g., due to the use of fixed keys, insecure cryptographic parameters and weak/broken cipher algorithms (e.g., RC4, DES).  \numc{299} apps transmit their insecure encrypted content over plain HTTP and non-HTTP protocols. %
In addition, we identified \numc{22} apps that perform their authentication over a secure channel (HTTPS) and then expose their authentication token over an insecure channel (HTTP and non-HTTP).
All these security issues enable a network adversary to read (even modify) sensitive plaintext information from encrypted traffic (e.g., using extracted fixed keys or breaking weak ciphers/keys).

\noindent(\textbf{5}) We also identify security and privacy issues beyond custom encrypted channels. For example,
we found that \numc{102} apps transmit their neighbor's wireless SSIDs to possibly track nearby users and their locations; \numc{202} apps collect/share the Android \textit{dummy0} interface information  (not protected by runtime/special permissions) that can be used for user tracking;
\numc{26} apps appear to allow UDP amplification, which can possibly be exploited in DDoS attacks.

\noindent(\textbf{6}) Besides app servers, tracking domains also receive various on-device information via non-standard channels. For example, \textit{appsflyer.com} may receive (depending on the app that includes the appsflayer SDK) items such as WiFi ESSID, WiFi MAC, operator, device email, build fingerprint, ad ID, and device ID, from 1386 of our tested apps with cumulative installations of over 24 billions. %

We will open source our tool at: \url{https://github.com/SajjadPourali/ThirdEye}.
We notified Google about the major privacy issues that we observed. We also contacted developers of 47 apps with significant security risks.

\begin{newtext}

\section{Threat model}
As we explore security issues due to the use of non-standard communication and custom encryption besides privacy exposure, here we also provide our threat model with different types of attackers, their capabilities, and goals. We exclude attacks that require compromising a user device or an app server. The attacks also do not involve other parties in the Android ecosystem such as device OEM providers, app developers, and app stores. The attacker cannot break modern crypto primitives, except when a key is exposed, or when a weak primitive is used—e.g., the attacker can brute-force a DES key, but not an AES-128 key (unless, e.g., a fixed AES key embedded in the app is used). The attacker can also monitor app behaviors on her own device (e.g., function hooking in a rooted phone), unless the app deploys active anti-analysis techniques that cannot be easily bypassed.

\subhead{On-path network attacker} This adversary has full access to the network communication between an app user and an app server, and can decrypt the encrypted content of network traffic, if insecure cryptographic keys (e.g., fixed keys extracted from an app), and weak algorithms (e.g., DES) are used. Such decryption will directly allow the adversary to access privacy-sensitive user information. The adversary may also get access to authentication tokens (if present) from such network traffic, which may lead to session hijacking and account takeover attacks.

\subhead{Co-located app attacker} This adversary has a regular app installed on the victim user’s device. With such co-located malicious apps, the attacker can access shared encrypted files saved by other apps on the same device, and decrypt such content, if insecure cryptographic keys or weak algorithms are used for encryption. This decryption may also expose a user’s private data.

\subhead{Device-owner attacker} In this case, we treat the device owner as the attacker, who would like to access protected (e.g., under custom encryption) service provider-content saved or processed on the device itself. This access may allow the attacker e.g.,  free access to VPN premium/paid services from the app provider.
\end{newtext}

\section{System design}
\begin{figure}[!htb]
\centering
  \includegraphics[scale=0.61]{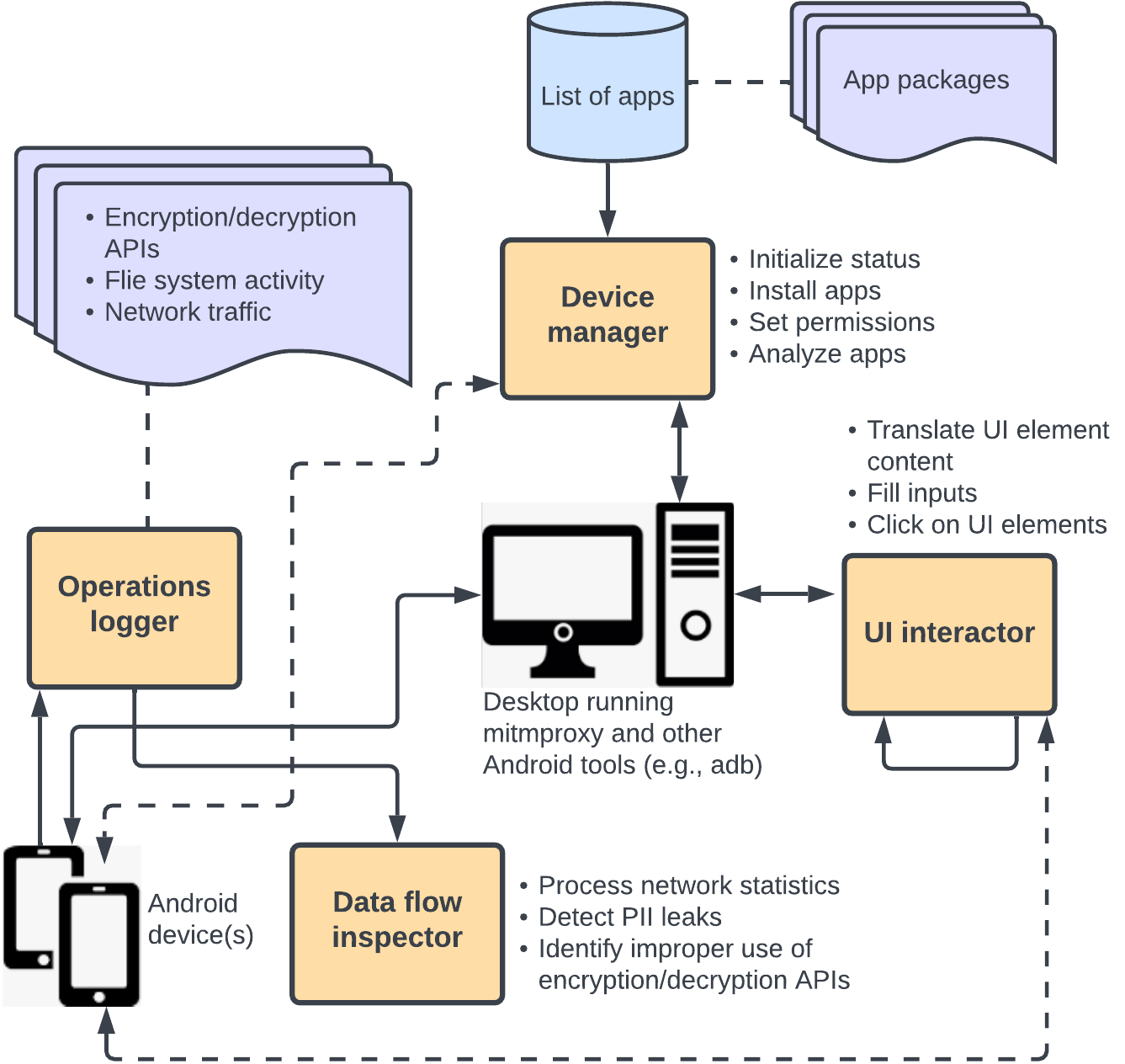}
  \vspace{-5pt}
  \caption{\ToolName\ design overview}
   \label{fig:system_design}
\end{figure}

In this section, we provide our design details; see Figure~\ref{fig:system_design} for an overview. To determine privacy and security issues resulting from non-standard and covert channels in apps,
we leverage network traffic captured from communication channels (HTTP/HTTPS and non-HTTP protocols), and cryptographic API parameters, and file operations during app execution. Our methodology requires rooted Android devices, and
consists of four main modules: the \textit{device manager} controls test devices and ensures that test prerequisites are satisfied; the \textit{UI interactor}
traverses and interacts with app menus to maximize code coverage; the \textit{operations logger}
locates cryptographic APIs, instruments cryptographic API parameters, and captures network traffic, extracts file operations; and the \textit{data flow inspector}
processes data flows to detect privacy and security issues.\looseness=-1

\subsection{Device manager}
As part of setting up the prerequisites, the device manager initiates a connection between the test desktop and Android devices through
ADB, and ensures that the devices are connected to the Internet. If the connection through ADB is successful, we uninstall all non-system apps (e.g., YouTube, Google Chrome) except our helper app that fixes the GPS location,
and prepare the device(s) for orchestration.

Given a candidate list of apps,
the device manager performs a cleanup (e.g., remove SD card content) before loading each app, removes the remnants left from the run of the previous app, downloads the corresponding app APK file from Google Play, or from alternative marketplaces (\url{apkpure.com} and \url{apktada.com}),
installs the app on the device,
sets all required runtime and special permissions, and proceeds with the analysis. Otherwise, the app is skipped.
Among special permissions, we consider only \textit{Accessibility}, and \textit{Notification Listener}; we exclude the ones requiring specific setup (e.g., VR permission), or the ones that can significantly affect UI/device operations (e.g., \textit{Display over other} or \textit{Device admin}).

The device manager then launches the app and monitors its progress. It can also detect and recover from possible failures (e.g., app closures, Internet outages).
It closes all installed apps to reduce the chance of UI misinteractions and capturing traffic originated from them (including traffic originated from the OS), but keeps running Android system services and other required apps (e.g., Android launcher) to keep the device functional.

To bypass commonly used runtime anti-reverse engineering protections and simulation of benign device conditions, we use seven modules: root-detection bypass, mock location detection, package installer detection, detection of the use of Frida, certificate pinning, ADB detection (detailed in Appendix~\ref{sec:runtime_anti_rev}).

\subsection{UI interactor} \label{sec:ui_interface_interactions}
During app execution, %
the UI interactor interacts with the app to increase the code coverage. %
We ensure that the target app is running in the foreground, and then explore and find different UI elements (including buttons, inputs, check-boxes) and interact with them (see Fig.~\ref{fig:ui_interaction} in the appendix).
To find inputs/clickable UI elements, we use a predefined keyword list (in English); see Table~\ref{tab:keyword_list} in the appendix.
To accommodate UI elements with non-English labels, we use the \textit{googletrans} Python module~\cite{googletrans} to translate the labels into English.
We then populate the input fields (if any) using a predefined list of inputs, and trigger the clickable UI elements based on the priority of a clickable element, which is determined based on its position in the list of keywords (e.g., the keyword of \textit{not now} %
has a higher priority than \textit{click}).
 After each click,
we add the clicked element to a list to avoid redundant actions. Clicking on an element may open new activities or trigger actions, and we follow the same steps for those new activities until all elements in the foreground app UI are explored. The \textit{back} button is clicked to go to the previous UI window (also used to avoid pop-up advertisements and in-app purchase windows).
We also identify and utilize the sign-up/sign-in functionalities to login to apps, e.g., by first using our Google test sign-in credentials in supported apps, and then creating an app-specific account (if possible); see Sec.~\ref{sec:interact_with_elements} for details.

\subsection{Operations logger} \label{sec:logger}

Apps may use socket APIs to communicate through non-HTTP channels in the transport layer and above (i.e., over TCP/UDP).
We use tcpdump~\cite{tcpdump} to store all network traffic in \textit{pcap} files, and thus, capture both HTTP and non-HTTP flows. We also log network tuples by hooking relevant APIs using Frida, to capture app specific network communication over sockets.
For detecting covert channels and misuses in shared storage, we hook \textit{open} and \textit{move} file API methods, to detect files that are opened/moved during an app's execution; we save these files for further analysis.
We use mitmproxy to capture/decrypt HTTPS traffic. The tcpdump data (not limited to HTTP/HTTPS) along with mitmproxy traffic obtained during network instrumentation is used to identify non-HTTP traffic. %

For cryptographic instrumentation, we capture (through Frida API hooking) API parameters used in cryptographic operations: plaintext, ciphertext, keys, initialization vectors (IVs), and cipher algorithms. To extract the parameters of Android SDK API, we hook \textit{init()}, \textit{update()} and \textit{doFinal()} API methods (of \textit{javax.crypto.Cipher API}~\cite{class_cipher}); note that \textit{getIV()} and \textit{getAlgorithm()} methods are called by the \textit{init() hook}. We define a non-SDK API as a third-party library used in an app, or a custom functionality implemented in an app that is not part of the Android SDK. %
Currently, we do not specifically handle obfuscated non-SDK APIs;
we look for \textit{encrypt} and \textit{decrypt} strings in method names to identify non-SDK APIs, and such keywords matching will not work with all obfuscated code.
If we find that an app uses encrypted/covert channels, we test the app on two separate devices, to identify fixed cryptographic keys used by the app.
We label a cryptographic key as fixed, when the same key value is returned from multiple runs of the instrumentation (on the same device and on different devices). %

\subsection{Data flow inspector}
This module detects privacy/security issues in non-standard and covert
channels in the collected network traffic.
We also leverage the collected parameters (i.e.,  ciphertext, plaintext, algorithm, key, IV) of encryption/decryption functions by hooking cryptographic API methods. Then we search the logged ciphertext from the captured content, and map/store the ciphertext with their corresponding plaintext.
We also check files stored on the device, including images, audio and video files. We categorize the captured content into HTTP, HTTPS, non-HTTP, and file.

For privacy issues, we extract Personal Identifiable Information (PII) and personal data (e.g., contacts, messages, images, audio, video) stored on the device to identify privacy exposures. We create copies of this data in different encoding formats (e.g., Base64, hex), and search these copies (exact and partial matches) within the network traffic, and magic headers (i.e., file signatures used to determine the content of a file) of transmitted media content (image, audio, video) originated from the device (i.e., outgoing).  Finally, we store the results of the search content in a database.%

For security issues, we check for situations where traffic sent over secure network channels are subsequently sent over insecure channels using HTTP/non-HTTP --- e.g., an authentication token sent over HTTPS, is subsequently sent over HTTP.
We also look for fixed/hard-coded keys in the app/library code, and the use of weak encryption algorithms (e.g., RC4) to encrypt data that is passed through insecure channels.

For covert channels, we check for files on shared storage that are opened by different apps. If we find files with common paths reused in multiple apps, we flag those as possible covert channels. %

\section{Implementation}
We use Python to implement \ToolName, and leverage the use of other Android command line utilities (e.g., ADB) to manage the orchestration of app executions. In addition, we use tcpdump and mitmproxy to capture network traffic and decrypt HTTPS communication.
We use Frida to instrument API methods,
implement the UI interactor component by extending AndroidViewClient.
We discuss our implementation details below. %

\subsection{Pre-execution steps}

To prepare an Android device for instrumentation, we first manually set the Android built-in WiFi proxy on the target device and import initial data on the device, including sample media files, contacts, and SMS messages. We also remove the device lock and increase display sleep timeout to avoid deadlocks in the UI interactor module.

We then use the \emph{app manager} component to handle downloading, installing and executing the latest and most compatible version (for our device hardware and OS) of a target app from Google Play. The app manager utilizes the UI interactor module to open and interact with Google Play, which is used to install apps; see Sec.~\ref{sec:Google-Play-store-integration}. The \textit{apkpure.com} and \textit{apktada.com} marketplaces are also checked if a target app is unavailable in Google Play (apps may fail to install from Google Play due to e.g., region locking).
During the first install of an app, we store all APK and OBB files, to avoid downloading them again for subsequent testing. %

The available permissions on a target device and the runtime permissions required to launch an app on the given device are extracted using the \textit{package manager (pm list permissions)}~\cite{pm} and \textit{dumpsys package <package>}~\cite{dumpsys} commands, respectively. We grant all the requested runtime permissions using the \textit{pm grant <package> <permission>} command.
Apps may also request special permissions (e.g., \textit{Accessibility}, \textit{Notification Listener}), which are only set via Android settings. We use the \textit{dumpsys package <package>} command to fetch services that request special permissions, and then execute the \textit{settings put} or \textit{cmd}~\cite{cmd} command (depending on the type of the requested permission) to grant the special permissions. %

\subsection{In-execution controller} %

We make sure that only our target app is installed on the device. Package names of all installed apps on the target device are extracted from \textit{cmd package list packages} command.  These package names are matched with that of the target app and allowed system apps (i.e., dependencies). Any apps with unmatched package names (i.e., non-system apps) are removed. Then,
prior to an app execution, we also ensure that all opened unwanted apps (e.g., Camera, Contact)
are closed using the \textit{pm clear} command~\cite{adb_clear}.
We also verify that the ADB connection between the test desktop and devices, and the Internet connection from the devices are successful.
We detect apps running in the background and foreground using the \textit{dumpsys activity activities} command~\cite{dumpsys}.
The output of this command returns structures\footnote{\url{https://android.googlesource.com/platform/frameworks/base/+/7efcc0c/services/java/com/android/server/am/ActivityStack.java}} showing foreground activity (\textit{mResumedActivity}) and background activity (\textit{mLastPausedActivity}) information. We make sure that the test app is always running in the foreground.

\begin{sloppypar}

Apps with certain permissions (e.g., CHANGE\_WIFI\_STATE) can perform disruptive operations (e.g., change WiFi connectivity state, screen rotation), which can affect our app analysis.\footnote{Although the \textit{setWifiEnabled} method was deprecated in Android 10, it still works in Android 12, for apps  built with a lower SDK API level --- see \url{https://developer.android.com/reference/android/net/wifi/WifiManager.html\#setWifiEnabled(boolean)}}
If an app disables WiFi, the Internet connection is lost, and if the screen rotates, a click event may trigger at the wrong position of the screen; to avoid such situations, \textit{svc wifi enable}~\cite{turn_on_off_wifi},
and \textit{content insert}~\cite{rotate_android_devices},
commands are used, respectively. Furthermore, because of the variation in the strength of the GPS signals received by a device, searching for the exact GPS location in the saved network traffic is problematic. The received GPS coordinates from satellites may vary slightly even when the device position is fixed.
Therefore, to return a fixed GPS value, we use our own GPS mocking app.

\end{sloppypar}

During UI interactions, it is possible to have accidental app closures or crashes.
Crashed or frozen apps are identified by inspecting the \textit{mCurrentFocus} structure that contains the current foreground window activity details. This structure is returned from the \textit{dumpsys activity activities} command. Therefore, \textit{mCurrentFocus} structure is inspected prior to UI interactions to detect crashes/freezes. The timestamp of the crash/freeze (if any) is also recorded. For crashed apps, we extract error logs using \textit{logcat}~\cite{logcat} (frozen apps do not produce any error logs). If an app crashes at startup, we try to rerun the app up to five times before skipping it. If the app crashes during execution, information collected so far is saved. %
When the app analysis completes, we save the analysis data on the device, if any (i.e., \textit{pcap} file, and files created by the apps). %

\subsection{User interface interaction}
We implement this module by extending the AndroidViewClient library that is designed to automate test scripts. AndroidViewClient provides UI based device interaction APIs (e.g., \textit{find}, \textit{click}, \textit{fill}). To find UI elements, it requires matching (exact/partial) keywords in a predefined list with UI element labels. Therefore, proper knowledge of the app view is required to determine what keyword list to use.

\subsubsection{UI element finder}
We use the \textit{dump} function in AndroidViewClient to get the foreground window content that contains all the window elements. If the element text in the UI window is non-English, the specific language is automatically detected and translated by googletrans~\cite{googletrans}.
To speed up the translation process, we store the translation result (i.e., original text and its English translation) in a database. We check this database before using googletrans for determining the translation of non-English text in the foreground window of the next app.

\subsubsection{UI element selection} \label{element_selection}
We create two separate lists for clickable and fillable elements.
The priority of selecting an element from these lists depends on the order of the elements in them (e.g., the keyword \textit{not now} has a higher priority than \textit{click}). The priority of an element depends on the order of appearance in the UI, e.g., \textit{accept}/\textit{submit} elements will appear after clicking \textit{agree}; see Table~\ref{tab:keyword_list} and Table~\ref{tab:input_list} in the appendix. We create the priority order list based on our observations from manually exploring several apps.
We then match the elements in the keyword list (based on the priority order) with the elements of app UI in foreground.

The clickable list contains the popularly used keywords in terms of clickability, with an optional exclude keyword list for each keyword to prevent interacting with similar words --- e.g., keyword \textit{agree} has an exclude list that contains \textit{agreement} and \text{disagree}. While \textit{agree} and \textit{agreement} are similar words, if an element with \textit{agree} is clicked, then clicking on a \text{disagree} element (an opposite action) is not possible.
The fillable list contains common keywords along with fillable values --- e.g., keyword \textit{email} with \textit{mymail@email.com} value (see Table~\ref{tab:input_list} in the appendix).

\begin{sloppypar}
To select clickable elements, we consider UI elements with the checkable/clickable property enabled, and have at least one of the following values in class attributes: \textit{android.widget.checkedtextview}, \textit{android.
view.view}, \textit{android.widget.button}, \textit{android.
widget.textview}, \textit{ndroid.widget.imageview} and \textit{android.widget.imagebutton}. To select fillable elements, we consider UI elements with the \textit{android.widget.edittext} value in class attributes.
\end{sloppypar}

\subsubsection{Interacting with UI elements} \label{sec:interact_with_elements}
Prior to interacting with UI elements, we wait for \numc{10} seconds to allow the target app to load. Then we find and fill all the fillable UI elements of the app UI (running in the foreground). If an app (e.g., a secure wallet app) prompts for the number pad, e.g., for a custom security PIN, %
we key in the digit 9 for 10 times, which we later search in the collected network traces and files (any fixed numeric sequence can be used).

\subhead{Identifying duplicate UI element visits}
To prevent duplicate visits to a UI element, we assign a unique ID to each element.
The ID is the concatenation of the element's \textit{view} attributes, and a \textit{window-hash}, defined as SHA256 of the \textit{dumpsys window | grep applicationId} command.
The element attributes (of the \textit{view}) that we leverage are: element ID, clickable, enabled. We order the element attributes prior to concatenating with \textit{windows-hash}.
Since the unique ID is composed by preserving the order of element interactions,
we prevent duplicate visits to UI elements and UI paths.

\begin{sloppypar}
\subhead{Identifying pop-up advertisements}
Prior to interacting with an app window, we check if it contains a pop-up advertisement; if so, the \textit{back} button is triggered to traverse to the previous app window. We currently consider ads served by the two most common (pop-up) ad platforms as we empirically observed in the top-100 Androidrank apps: \textit{Google AdMob}~\cite{Admob} for non-gaming apps and \textit{Unity Ad Units}~\cite{AdUnits} for gaming apps.
We detect AdMob pop-up ads using
the \textit{com.google.android.gms.ads.AdActivity} activity, and Unity pop-up ads using
\textit{com.unity3d.ads.adunit.AdUnitActivity} and \textit{com.unity3d.services.ads.adunit.AdUnitActivity} activities. The support for other ad management SDKs can be easily added.

\end{sloppypar}

\begin{sloppypar}
\subhead{Identifying Google in-app purchases}
In-app purchase features can negatively affect the analysis by deviating the UI interactor to deal with third-party components, instead of the app itself. To address this issue, we use \textit{dumpsys activity} to detect the Google Play in-app billing (in-app purchase) activities. We identified  the activities that belong to the Google's in-app purchase windows.\footnote{The activities are: \textit{com.google.android.finsky.activities.MarketDeepLinkHandlerActivity}, \textit{com.google.android.finsky.billing.acquire.LockToPortraitUiBuilderHostActivity}, and
 \textit{com.google.android.finsky.billing.acquire.SheetUiBuilderHostActivity}.} Therefore, we trigger the \textit{back} button to go to the previous window, if we encounter a Google in-app billing window.

\end{sloppypar}

\subhead{Google sign-in authentication}
If the \textit{google} keyword appears in the clickable UI elements,
it usually indicates the app's support for Google sign-in; we also check for relevant activities.\footnote{The activities are: \textit{com.google.android.gms/signin.activity.ConsentActivity} and \textit{com.google.android.gms/auth.uiflows.consent.BrowserConsentActivity}.} We then use the email address registered with the device to authenticate. The UI interactor also grants access for sign-in activities and relevant permissions by clicking the \text{confirm} button (if prompted).

\subhead{Terminating UI interaction}
To prevent the exhaustion of system resources, \textit{\ToolName} interacts with an app until one of the following conditions is met: (i) the number of interaction attempts with UI elements reaches 100; (ii) no new elements are found; (iii) the app cannot be opened
even after 5 consecutive attempts; (iv)
the duration of the interactions reaches 5 minutes. %

\subsubsection{Google Play Store integration}
\label{sec:Google-Play-store-integration}
We use Google Play as the default app market. The target app's installation window is opened by using the Android Intent-filter, \url{market://details?id=PKGNAME}.
Then the UI interactor (see Sec.~\ref{sec:ui_interface_interactions}) installs the app from Google Play, by clicking the \textit{Install} button (if available).
We check every 10 seconds for the presence of an \textit{open} button, to confirm that the app is successfully installed; after 200 seconds, the installation is skipped. To deal with common app installation prompts (e.g., agreeing to install, consenting to permissions, providing credit card information), we handle various UI elements with labels including: \textit{try again}, \textit{retry}, \textit{accept}, \textit{update}, \textit{skip}, \textit{accept}, \textit{no thanks}, \textit{continue}, and \textit{ok}.\looseness=-1
\subsection{Instrumentation}
We describe the following instrumentation methods used in \ToolName. %
We complement the use of Frida to comprehensively collect the instrumented data.

\subhead{Network and file instrumentations}
We use tcpdump to collect non-HTTP traffic, and mitmproxy to capture HTTP/HTTPS traffic.
We use the global proxy settings of Android devices to forward the HTTP/HTTPS traffic to an mitmproxy server running on the test desktop. As some apps ignore the proxy setting,
we hook (via Frida) the remote TCP connections with port 443 that bypass the global proxy, and forward the traffic to our desktop mitmproxy. %
For files, we hook \textit{open}, \textit{remove}, \textit{rename}, \textit{read} and \textit{write} Bionic library functions,
which are used for shared storage operations. These functions cover file operations used in both Android SDK and NDK. We store read and write buffers, and process them later.

\subhead{Rule-based API hooking}
We implement a rule-based hooking module using Androguard
and Frida.
This module provides the ability to define selection criteria and actions on API methods in DEX files to choose and trigger dynamic actions (e.g., logging or changing parameters) by accepting callback functions. We use Androguard to select methods based on defined criteria and Frida to perform the defined action. Androguard is used to fetch all the declared API methods in the DEX files that use \textit{EncodedMethod} (an Androguard Object), which contains the method name, parameters, return type, class name, Dalvik bytecode (of the method). Since Androguard works with \textit{Dalvik} method definition syntax, and Frida uses Java method definition syntax, our module maps Androguard results to Java format. Then we create a hooking script for Frida, based on the defined callback functions that would be executed by Frida when called.
We primarily use this module to evade root detection and to log non-SDK encryption/decryption methods; however, it is generic enough to be reused for other purposes. %

\subhead{Cryptographic instrumentation}
 To collect cryptographic parameters, we log the input parameters, return value, execution timestamp and object ID of each method.
 For this purpose, we hook \textit{init()} (for the parameters such as the key, IV, algorithm, and operation type), and \textit{doFinal()} and \textit{update()} (for plaintext and ciphertext) Android SDK cryptographic API methods from \textit{javax.crypto.Cipher}.
 To relate these API calls in sequence, %
 we use their object IDs and execution timestamp. \new{Note that besides \emph{decrypt} functions, we can also collect necessary plaintext items only from \emph{encrypt} functions---i.e., we log the inputs before they are encrypted, and thus we are unaffected by apps' not invoking \emph{decrypt} functions.} %

Android SDK cryptographic APIs cover both single-part and multi-part encryption/decryption. Multi-part operations are usually used when the data is not contiguous in memory, e.g., large files, socket streams. To defragment multi-part blocks, we trace back \textit{update()} and \textit{doFinal()} functions based on their object hashcode~\cite{java7_api_spec}
and calling timestamp, until a \textit{javax.crypto.Cipher} object initialization or a cryptographic parameter modification occurs.

We also look for non-SDK cryptographic APIs in apps. We leverage our rule based logger to find all methods with \textit{encrypt} and \textit{decrypt} in their method names, which at least accept one argument in byte or string types, and return the byte or string type. After identifying the specific APIs methods, we automate the creation of corresponding Frida API method hooks, and log their arguments and return values. In addition, we observe the logged arguments to detect potential cryptographic keys by looking for arguments that are of \numc{128}, \numc{192}, or \numc{256} bits in length, which come with other arguments that have any length except \numc{128}, \numc{192}, or \numc{256} bits.

 We identify nested encryption/decryption by recursively checking ciphertext for the corresponding plaintext in the collected instrumented data. For each level of nested encryption, we create a new encryption entity with corresponding parameters. If the nested plaintext is compressed, we also consider its decompressed value.

\begin{sloppypar}
\subhead{Android ID, PII and device Info}
We manually run \textit{getprop}, \textit{ifconfig} and \textit{dumpsys} commands (using ADB) to extract all available persistent PII and device information in JSON format except for three identifiers, which are not persistent --- i.e., \textit{Android ID}, \textit{Advertising ID} and \textit{Dummy0 Address} that are automatically extracted during app interaction by calling \textit{getAdvertisingIdInfo} API, \textit{Settings.Secure{\#}ANDROID\_ID} API and \textit{ifconfig} command (with ADB), respectively. Note that apps installed on devices using Android 8.0 and above get a unique Android ID value for each app, which is composed of the app sign-in key, device user ID and device name. The non-persistent data is individually stored for each analyzed app, allowing us to perform multi-device analysis by choosing the appropriate PII items collected during our inspection.%
\end{sloppypar}

\subsection{Inspection}\label{sec:inspection}

We categorize the collected network communication and file operations of each app: HTTP, HTTPS, non-HTTP, file. Then we store the details of the instrumented data (i.e., destination, direction to/from the device, headers of HTTP/HTTPS traffic and content) in separate lists maintained for each category. Before storing the information in the lists, we use \textit{python-magic}~\cite{python_magic} to identify and decompress the compressed data (if any). We then search PII data on these lists.

\subhead{Non-HTTP communication}
To extract non-HTTP communication, we remove system traffic from the captured pcap file, then we parse it using dpkt~\cite{dpkt}, to determine if the corresponding TCP/UDP packets are non-HTTP. The application layer protocols for non-HTTP traffic do not include TLS or HTTP/S.%

\subhead{Defragmentation}
Fragmentation can affect our inspection of network traffic because the standard MTU~\cite{what_is_mtu}
of IP datagrams over Ethernet is \numc{1500} bytes (same as in the WiFi interface~\cite{rfc0894}). Therefore, any IP datagram over \numc{1500} bytes will be fragmented. As a result, we will not find PII values (if exist) that are split between multiple packets. To overcome this problem, we defragment to recover the original data of the fragmented TCP packets,
and use the dpkt~\cite{dpkt} library to parse TCP and UDP traffic data.

\subhead{Identifying encrypted data}
We extract ciphertext values from cryptographic APIs (e.g., \textit{Cipher}) and search them in lists created for all categories (i.e., HTTP, HTTPS, non-HTTP, file). If a ciphertext value is found in the content of any of the lists, we add  its cryptographic parameters to a new list with the same name and an additional \textit{encrypted} suffix. Apps can send the ciphertext in chunks. Therefore, we split the ciphertext into 18 bytes chunks (assuming 128-bit blocks), which reduces the chance of getting identical blocks by covering more than one block of a cipher text, prior to searching them in the lists pertaining to different categories. %

\subhead{Search strategy}
The content in the network traffic can be transformed into different forms.
It is also possible that one (e.g., capitalize, upper case, lower case, Base64) or more (e.g., \textit{md5-hex} --- creates an MD5 hash with hex encoding, \textit{sha1-hex}, \textit{sha256-hex}, \textit{md5-raw}, \textit{sha1-raw}, \textit{sha256-raw}) of these transformations are applied to the content.
Therefore, we compile a set of values (e.g., PII, list of keywords, cryptographic keys and fillable content; see Table~\ref{tab:input_list} in the appendix), and apply the mentioned transformations to each value in the set, and save them in a separate list, which we then use to search and identify privacy/security issues.

\subhead{Detecting insecure cryptographic parameters}
We use \textit{apktool}~\cite{apktool} to unpack APK files, and search the collected keys in different encoding formats (plain, Base64, hex case-insensitive) over all the unpacked content of APK files, to determine if any of the fixed keys are hard-coded (see Sec.~\ref{sec:logger}).
Thereafter, we collect the keys of the traced ciphertexts in the network communication or files.
If we detect hard-coded/fixed (i.e., reused) keys from the network communication in multiple runs, and on the same or different devices, we mark them as insecure keys.  %

\subhead{App and system traffic separation}
The captured traffic from tcpdump and mitmproxy may contain traffic from system processes running on a device, which is separate from the app traffic. To ensure that we only analyze the traffic of the target app, we filter the captured network packets using the collected network tuples by API hooking and their timestamps (see Table~\ref{tab:instrumentation_network_api} in appendix). We hook the process ID of the target app, to ensure system/app traffic separation --- all hooks are at the app level. %

\section{Results}
In this section, we summarize our findings on privacy and security issues of Android apps that use non-standard and covert  communication channels. %
\new{
Instead of choosing top-downloaded apps, which may not cover various app categories, we selected apps from Androidrank~\cite{androidrank} for our evaluation. Androidrank ranks Google Play apps in 33 categories based on various metrics such as total downloads, total number of user ratings, average user ratings. We collected all available \numc{15522} %
unique free apps for our evaluation from all categories (note that there are overlaps between app categories). This dataset contains apps that are highly popular globally (e.g., 1B+ installs), but also apps that are top-ranked (within top-500) in a specialized app category with a relatively small number of installations (e.g., 10K+); see Fig.~\ref{fig:comp_pct_apps_installs} in the appendix.
}
\textit{\ToolName} could download \numc{15327} apps,
and successfully analyzed \numc{12598} apps; the remaining \numc{2729} apps failed for various reasons, e.g., app incompatibility with Android 12, geo-blocking, unknown reverse engineering protection, and app-crashing due to the use of Frida method hooking.
We ran our experiments between Nov.\ 25, 2021--Jan.\ 6, 2022.
We used two Android devices (Pixel 4 and Pixel 6) running factory images with Android 12, and a desktop running Ubuntu 21.04, Core i9-10900, 64GB RAM, 2TB storage. Most apps finished their execution (i.e., all their UI interactions were completed) within 5 minutes; we terminated the execution of \printpercentoutof{1329}{12598} apps at the 5-minute threshold (see Fig.~\ref{fig:app_interaction_timings} in the appendix).
For a summary of our results, see Tables~\ref{tab:on-device-info} and \ref{tab:content_types_on_channels}. We also provide several  examples of prominent privacy/security issues from our findings in Sec.~\ref{sec:discussion}. \new{We report some additional network security results in Appendix~\ref{sec:netsec-misc}.}

We categorize privacy-sensitive data into \textit{Device}, \textit{Network}, \textit{Network Location}, \textit{GPS Location}, and \textit{User} categories; see  Table~\ref{tab:on-device-info}.
We also label the likely use of the available data into \textit{Persistent ID}, \textit{Short-term}, \textit{Profiling}, \textit{Location Data},  and \textit{User Assets}. Items labeled as \textit{Persistent ID} and \textit{Short-term} are generally used for tracking; \textit{Persistent IDs} do not change with time, and \textit{Short-term} items can identify a user for a short duration (can be used for long-term tracking if combined with other items). Profiling items can identify a user, or a user-group to a varying degree, the accuracy of which improves when combined.

\newcommand{\STAB}[1]{\begin{tabular}{@{}c@{}}#1\end{tabular}}
\begin{table*}[htb]
    \centering
    \resizebox{\linewidth}{!}{
\begin{tabular}{ @{}c@{ } l l l |r r |r r |r r |r r r r}
 \multirow{2}{*}[0ex]{\small \textbf{Category}}&\multirow{2}{*}[0ex]{\small \textbf{Data Type}}&\multirow{2}{*}[0ex]{\small \textbf{Protection level}}&\multicolumn{1}{l}{\multirow{2}{*}[0ex]{\small \textbf{Purpose/Use}}} &\multicolumn{2}{c}{\textbf{HTTPS}}&\multicolumn{2}{c}{\textbf{HTTP}}&\multicolumn{2}{c}{\textbf{Non--HTTP}}&\multicolumn{3}{c}{\textbf{Network-wide}} \\
&&&&{\cellcolor{gray!25}\small Regular}&{\cellcolor{gray!25}\small \shortstack{Custom\\Encrypted}}&{\cellcolor{gray!25}\small Regular}&{\cellcolor{gray!25}\small \shortstack{Custom\\Encrypted}}&{\cellcolor{gray!25}\small Regular}&{\cellcolor{gray!25}\small \shortstack{Custom\\Encrypted}}&{\cellcolor{gray!25}\small Regular}&{\cellcolor{gray!25}\small \shortstack{Custom\\Encrypted}}&{\cellcolor{gray!25}\small Overall}\\
\hline
\multirow{11}{*}{\STAB{\rotatebox[origin=c]{90}{{\small Device}}}}
&Device ID&Normal&Persistent ID$^\ddagger$&4504&526&347&100&30&10&4678&595&4818\\
&Advertising ID&Normal&Persistent ID$^\dagger$&7812&1990&312&100&5&3&7841&2034&8006\\
&Bootloader&Normal&Profiling&131&27&2&0&1&0&133&27&160\\
&Build Fingerprint&Normal&Profiling&474&51&8&8&3&0&482&58&532\\
&CPU Model&Normal&Profiling&795&128&22&3&4&0&814&131&919\\
&Display ID&Normal&Profiling&10376&1705&1925&18&8&0&10725&1712&10726\\
&Device Name&Normal&Profiling&3960&1605&190&14&20&6&4057&1614&4918\\
&Device Resolution&Normal&Profiling&489&29&28&4&0&0&515&33&540\\
&Device ABI&Normal&Profiling&1754&1548&26&11&6&0&1552&1773&2971\\
&Device Model&Normal&Profiling&12031&1966&2029&92&93&8&12289&2007&12289\\
&Dummy0 Interface&Normal&Short-term&183&8&5&3&0&3&187&14&201\\
\hline
\multirow{6}{*}{\STAB{\rotatebox[origin=c]{90}{{\small Network}}}}
&Operator&Normal&Profiling&5253&1595&106&17&12&0&5274&1607&5563\\
&Device WiFi IP&Normal&Short-term&1030&172&23&12&21&2&1060&179&1210\\
&Device WiFi IP6&Normal&Short-term&64&10&2&1&0&0&65&11&75\\
&Device Proxy IP&Normal&Short-term&36&11&2&5&0&1&38&17&53\\
&Default Gateway IP&Normal&Short-term&736&139&27&11&13&3&764&149&888\\
&WiFi MAC&Normal&Persistent ID$^\dagger$&63&9&19&9&2&3&80&17&88\\
\hline
\multirow{4}{*}{\STAB{\rotatebox[origin=c]{90}{\parbox{1cm}{\small\centering Network\\Location}}}}
&Router ESSID&Dangerous&Location Data&216&39&4&8&0&5&218&51&260\\
&Router BSSID&Dangerous&Location Data&207&37&19&4&0&5&217&46&255\\
&neighbor Router ESSID&Dangerous&Location Data&74&15&2&2&0&0&75&17&91\\
&neighbor Router BSSID&Dangerous&Location Data&61&22&0&1&0&0&61&13&74\\
\hline
\multirow{3}{*}{\STAB{\rotatebox[origin=c]{90}{\parbox{1.1cm}{\small\centering GPS\\Location}}}}
&GPS ($\leq$7 meter accuracy) &Dangerous&Location Data&1352&68&65&14&1&0&1397&80&1448\\
&GPS (78 meter accuracy) &Dangerous&Location Data&1637&71&81&15&1&0&1687&84&1738\\
&GPS (787 meter accuracy) &Dangerous&Location Data&1742&74&84&16&1&0&1792&88&1844\\
\hline
\multirow{6}{*}{\STAB{\rotatebox[origin=c]{90}{{\small User Assets}}}}
&List of Apps&Normal&Profiling&38&15&4&5&1&0&41&20&61\\
&SMS&Dangerous&User Asset&1&0&0&0&0&0&1&0&1\\
&Phone Number&Dangerous&Persistent ID&26&5&1&1&0&0&27&6&32\\
&Contacts&Dangerous&User Asset&7&1&0&0&0&0&7&1&8\\
&Device Email&Dangerous&Persistent ID&924&42&21&3&1&0&941&44&966\\
&User Files&Dangerous&User Asset&-&45&-&7&-&0&-&52&52\\\hline
\end{tabular}
}
    \caption{Transmission of the on-device information over the network -- without or with custom encryption, listed under the \emph{Regular} and \emph{Encrypted} columns, respectively. Items marked with $^\dagger$ can be fixed or reset/changed by the user or system choice; $^\ddagger$ marked items are considered persistent up to Android version 8 (app-specific afterward). The last column (Overall) represents all the apps that use regular or custom encrypted channels, excluding the apps that use both channels for the same data type.}%
    \label{tab:on-device-info}
    \vspace{-10pt}
\end{table*}

\subsection{\mbox{Characteristics of encrypted communication}}
\label{Sec:Result-Characteristics}

\subhead{Prevalence of the use of encryption}
We found that \printpercentoutof{6075}{12598} apps triggered encryption/decryption related calls from our Frida API hooking. From these apps, we identified \printpercentoutof{2887}{6075} apps send network traffic, and %
use file storage with data originating from the hooked encryption/decryption calls; the remaining apps possibly use such calls for internal/local purposes.
We found 4 apps that used two nested layers of encryption, although no relevant traffic was observed during our test window;  e.g., \textit{com.mci.balagh} (Ministry of Commerce of Saudi Arabia) app, hard-coded its remote server address in an encrypted form (nested), and subsequently decrypted twice. %
In terms of encryption type, we observed \numc{2597}, \numc{598}, \numc{119} apps used symmetric, public key, non-SDK encryption,
respectively;
see Table~\ref{tab:prevalance_enc} (in the appendix).

\subhead{Encrypted communication content} To identify the type of content sent over encrypted channels, we created a list of keywords (see the \textit{Data Type} column in Table~\ref{tab:on-device-info}):
device information used for tracking (e.g., network operator, build fingerprint), network information (e.g., device MAC), GPS coordinates in different accuracies, network location (e.g., via own/neighbor router info), and user assets (e.g., contact list, SMS). We also extract authentication tokens and session IDs embedded
in JSON, XML, HTTP headers, form-urlencoded, and form-data data structures, besides authentication passwords (see the \textit{User Credentials} column in Table~\ref{tab:content_types_on_channels}). We did not verify the tokens used for \textit{User Credentials} (except a few selected ones for manual verification, e.g., \textit{com.peppermint.livechat.findbeauty}).
Apps also exchange symmetric encryption keys over HTTP/HTTPS and non-standard channels: 82 apps sent and 10 apps received such keys over HTTP; 154 apps sent and 71 received such keys over HTTPS; and 8 apps sent such keys over non-HTTP.

\subhead{Encrypted communication channels}
To understand information leakage between different transmission channels, we categorize such channels into the following four categories.
We consider that an app transmits a leaked item (e.g., Device ID) through a \textit{Regular} channel, if the app shares the item using HTTP/S; the app may also apply custom encryption for this transmission (e.g., to the same or different hosts). For \textit{Custom Encrypted}, the leaked item is shared via at least one channel after processing the item with one or more additional encryption layers; the same item may also be shared via \textit{Regular} channels. We use \textit{Custom Encrypted for Some Hosts} for apps that share the leaked item with one or more distinct remote hosts, only under custom encryption; this leakage will be missed by other tools (although the same information leakage will be detected for other hosts when shared via \emph{Regular} channels).
If an app uses only custom encrypted channels for sharing the leaked item, which is not shared via \emph{Regular} channels, we count such app under \textit{Only Custom Encrypted}; existing tools cannot detect any leakage from this category.
See Table~\ref{tab:content_types_on_channels} for overall results for these channels, and Sec.~\ref{sec:discussion} for prominent examples.

\begin{sloppypar}
\subhead{Recipients of encrypted traffic} %
\numc{1291} and \numc{786} unique remote servers with registered domain names and subdomain names, respectively, were contacted by the \numc{2887} apps that used additional layers of encryption. See Table~\ref{tab:encrypted_destinations} (in the appendix) for the top-10 remote servers (all tracking SDKs) receiving various on-device information. Some destinations may receive several on-device information items (e.g., \textit{appsflyer} receives items such as WiFi ESSID, WiFi MAC, operator, device email, build fingerprint, advertisement ID, device ID), and other destinations may receive very basic items (e.g., \textit{scorecardresearch.com} receive only advertisement ID). %
More importantly,
22 apps sent users' GPS coordinates to these domains:
\numc{10} apps to \textit{appsflyer.com} (3 only under custom encryption), \numc{8} apps to \textit{supersonicads.com}, \numc{3} apps to \textit{batch.com} (2 only under custom encryption), and one app to \textit{pangle.io}.
\textit{Appsflyer.com} also received search terms from two applications (\textit{ru.labirint.android} and \textit{com.lotte}), and the user-entered phone number from another app (\textit{vn.gapo.app}).
\end{sloppypar}

\subhead{Encrypted channels with packers}
Android apps can use packers to protect apps from being copied or altered (e.g., by encrypting class DEX files). We used \textit{APKiD}\footnote{\textit{APKiD} (\url{https://githubhelp.com/l9sk/APKiD}) provides information on how an APK is formulated, e.g., compilers, packers, obfuscators.}
to identify the prevalence of packers in apps. We found that \printpercentoutof{121}{12598} apps use packers for Java implementations. These apps can contain some API methods not detectable by common static analysis tools (e.g., \textit{Apktool} and Androguard). \textit{\ToolName} uncovered \printpercentoutof{51}{121} apps that use cryptographic APIs, and \printpercentoutof{34}{51} apps that use custom-encrypted channels leveraging packers.
In addition, by analyzing 20 randomly selected apps that use \textit{appsflyer} tracking SDK, we found all of those apps included packed \textit{appsflyer} SDK, but \textit{APKiD} failed to identify the packed SDK. This SDK was used in \printpercentoutof{1386}{2887} apps that used custom-encrypted channels to send tracking information (see under  ``Recipients of encrypted traffic'').

\begin{table*}[!htb]
\centering
\resizebox{\linewidth}{!}{
\begin{tabular}{ l l r r r r r r r r r r r}
\multirow{2}{*}[0ex]{\textbf{Protocol}} & \multirow{2}{*}[0ex]{\textbf{Channel}} & \multirow{2}{*}[0ex]{\textbf{Device}}& \multirow{2}{*}[0ex]{\textbf{Network}} &\multicolumn{2}{c}{\textbf{Network Location}}&\multirow{2}{*}[0ex]{\begin{tabular}{@{}c@{}}\textbf{GPS} \\ \textbf{Location}\end{tabular}} & \multirow{2}{*}{\begin{tabular}{@{}c@{}}\textbf{User} \\ \textbf{Assets}\end{tabular}}& \multicolumn{2}{c}{\textbf{Credentials}}  &
\multirow{2}{*}{\begin{tabular}{@{}c@{}}\textbf{Key} \\ \textbf{Transmission}\end{tabular}}\\
&&&&\cellcolor{gray!25}Own Router&\cellcolor{gray!25}Neighbor Router&&&\cellcolor{gray!25}Password&\cellcolor{gray!25}Token&\\ %
\hline
\multirow{4}{*}{HTTP} & Regular &2109&150&20&2&197&26&8&157&0\\
 & Custom Encrypted &191&34&8&2&17&15&0&20&87\\
 & Custom Encrypted for Some Hosts &93&32&7&2&13&13&0&17&86\\
 & {\bf Only Custom Encrypted} &\textbf{36}&\textbf{17}&\textbf{7}&\textbf{2}&\textbf{10}&\textbf{13}&\textbf{0}&\textbf{15}&\textbf{86}\\
\hline
\multirow{4}{*}{HTTPS} & Regular &12178&5640&256&80&2442&985&327&9019&0\\
 & Custom Encrypted &2272&1686&49&20&87&104&15&378&182\\
 & Custom Encrypted for Some Hosts &1953&1663&45&20&83&97&15&316&181\\
 & {\bf Only Custom Encrypted} &\textbf{1443}&\textbf{429}&\textbf{39}&\textbf{19}&\textbf{46}&\textbf{85}&\textbf{15}&\textbf{221}&\textbf{181}\\
\hline
\multirow{4}{*}{Non-HTTP} &Regular &120&26&0&0&1&2&0&0&0\\
 & Custom Encrypted&10&5&5&0&0&0&0&4&8\\
 & Custom Encrypted for Some Hosts&10&5&5&0&0&0&0&4&8\\
 & {\bf Only Custom Encrypted} &\textbf{3}&\textbf{1}&\textbf{5}&\textbf{0}&\textbf{0}&\textbf{0}&\textbf{0}&\textbf{1}&\textbf{8}\\
\hline%
\multirow{4}{*}{Overall} & Regular&12420&5690&266&81&2576&1006&334&9063&0\\
 & Custom Encrypted&2350&1707&61&22&102&117&15&398&263\\
 & Custom Encrypted for Some Hosts &1996&1681&58&22&95&109&15&337&263\\
 & {\bf Only Custom Encrypted} &\textbf{1481}&\textbf{451}&\textbf{51}&\textbf{21}&\textbf{56}&\textbf{97}&\textbf{15}&\textbf{237}&\textbf{263}\\
\hline
\end{tabular}}
    \caption{Content types sent over different protocols and channels. For channel categories, see ``Encrypted communication channel'' in Sec.~\ref{Sec:Result-Characteristics}.} %
    \label{tab:content_types_on_channels}
\vspace{-10pt}
\end{table*}

\subsection{Insecure key management and weak ciphers}

We found \printpercentoutof{2421}{2887} apps sent data with custom encryption using fixed keys
(on the same device in two different installations); \printpercentoutof{2112}{2421} apps used symmetric and \printpercentoutof{502}{2421} apps used public-key ciphers. %
On the other hand, \printpercentoutof{1780}{2421} apps used fixed keys across devices; \printpercentoutof{1593}{1780} and \printpercentoutof{341}{1780} of them used symmetric and public-key ciphers, respectively.
Moreover, we identified \printpercentoutof{561}{2421} apps with hard-coded keys. %
We also identified \printpercentoutof{154}{2421} apps used both symmetric and public-key ciphers with fixed keys.
We also observed that \numc{27} apps used custom encryption to store their content in the device shared storage; \numc{26} apps used symmetric keys, and one used a public key; \numc{9} apps stored ciphertext (generated using symmetric fixed keys) in shared storage, %
exposing
various content types (e.g., device information, inputs, network data) to other apps; see Table~\ref{tab:onsecure_key_mgmt} (in the appendix) and Sec.~\ref{Sec:Privacy_exposures_from_files}.

In terms of the use of broken/weak cryptographic algorithms, and short-length keys, we observed that even Android 12's cryptographic API does not restrict such usage; see Table~\ref{tab:weak_cipher_protocols} (in appendix). We identified \printpercentoutof{262}{2887} apps used insecure algorithms, e.g., DES (\numc{106}), RC4 (\numc{3}), 3DES (\numc{34}), RSA-384 (\numc{1}), and RSA-512 (\numc{60}). %
The use of fixed keys and weak ciphers can lead to serious privacy/security issues, depending on the app; see Sec.~\ref{sec:discussion} under ``New security vulnerabilities''. Note that if an app uses a fixed/hard-coded key to encrypt data sent over HTTPS, then this will not lead to data exposure to a network attacker.

\subsection{Apps sending geolocation information}
\subhead{GPS and router SSID}
We observed that \printpercentoutof{2727}{12598} apps sent GPS coordinates~\cite{wiki:Decimal_degrees} and router's SSID to remote servers; \printpercent{129}{2727} of them used additional encryption to send this information; see Table~\ref{tab:on-device-info}.
Interestingly, \numc{197} apps sent GPS coordinates to third-party services, but not to their own servers.
For example, the official app of Russian Post (\textit{com.octopod.russianpost.client.android}) sent GPS coordinates (via regular HTTPS) to  \textit{tracker-api.my.com}, a subsidiary of the Russian social media company VK (\url{vk.vom}). %
On the other hand, \textit{com.cashingltd.cashing}, \textit{com.tdcm.android.trueyou} sent GPS coordinates to \textit{appsflyer.com} only under custom encryption.

\subhead{Neighbor's router scanning}
Apps with location permission can collect BSSID, ESSID from the app user's router, as well as all nearby wireless routers. Such router IDs have been used to determine physical location since 2011 (e.g.,~\cite{bssid-google-2011}), and currently public databases of such ID-location mapping exist for hundreds of millions of routers (see e.g., \url{wigle.net}, \url{mylnikov.org}); this has also been exacerbated by the increasing adoption of IPv6~\cite{bssid-bh2021}. A user's location-capable app thus can reveal not only her location, but also the location of her neighbors (irrespective of the apps/devices used by them). We found \numc{102} apps that sent neighboring router IDs to their servers (notable apps: PayPal, PayPal Business, Yandex, Mail.ru, VK, Kaspersky Security and VPN). More importantly, \printpercent{22}{102} apps sent such IDs only via custom encrypted channels; a notable example: Baidu Search (\textit{com.baidu.searchbox}). Even after a user moves to a new location with her old router, her location change can still be exposed, if she has a neighbor with a location-capable app. The \numc{102} apps that we identified, have been mostly downloaded by users from Russia (\numc{66480721} users), Brazil (\numc{41163244}), Indonesia (\numc{9566304}), USA (\numc{8802562}), and India (\numc{6749443}); estimated download numbers are from \textit{similarweb}~\cite{similarweb} (Q2, 2021, for Google Play apps). Some of these non-Google-Play apps are also very popular; e.g., \textit{com.baidu.searchbox} and \textit{com.sina.weibo}, ranked 9th and 12th, respectively, in \url{AppinChina.co} app store.

\subsection{Exposures via files} %
\label{Sec:Privacy_exposures_from_files}

\subhead{Leftover files in external storage}
Among  our analyzed apps that created files in external storage, 128 apps stored device information, 12 stored GPS coordinates, and 10 stored network information.
\printpercentoutof{27}{150} of these apps used \TechniqueName\  to store content in external storage; \printpercentoutof{9}{27} apps stored device info and one of those apps stored authentication tokens; e.g., \textit{ru.mediafort.povarenok} stored the DES-encrypted value of the device email (i.e., device account) in \textit{mediafort/data.txt}; \textit{tw.comico} stored user authentication tokens with a fixed key.

\subhead{Covert channels}
We found \numc{44} apps stored device information into common folder paths in shared storage. There were \numc{104} apps that checked the existence of these paths. These files can be used as inter-process communication (IPC)/covert channels --- \numc{4} apps from different vendors wrote the device WiFi MAC address to \textit{.cc/.adfwe.dat} file path; \numc{8} apps from different vendors check the existence of this path; \numc{20} apps saved the MD5 hashed value of WiFi interface MAC address; \numc{67} apps check the existence of the path. Moreover, we detect that \textit{app.buzz.share} app and its lite version with over \numc{110} million downloads stored identifiers, such as device ID, in a file (\textit{bytedance/device\_parameters\_i18n.dat}), encrypted with DES.
Three more apps from different vendors saved the same data, key, and encryption algorithm information to the same path.

\begin{newtext}
\section{Effectiveness and Limitations}

\subhead{Overall effectiveness}
We verified our initial results through manual inspection, refined the tool before conducting the large-scale study. Note that we do not have any ground-truth on the targeted leakage, and we also cannot rely on any existing tools for accuracy; e.g., AGRIGENTO~\cite{obfus_res} could have been used in some limited cases (e.g., for the data types considered by the tool), but it is now outdated (designed for Android 4).
\ToolName's effectiveness is apparent from the numerous new privacy exposures of various types that we uncovered. %
However, for some apps, our analysis may fail to fully identify the security and privacy issues due to the use of non-standard and custom encryption channels---see below under limitations. We first summarize the strengths of \ToolName\ components, which, in combination, contribute to our overall effectiveness.

Our UI interactor (partially) supports custom registration/login and Google sign-in, detects already explored widgets/objects to prevent duplicate interaction/exploration, and avoids non-targeted activities (e.g., ads). In contrast, Android Monkey lacks these features, and hence can take longer for the same code coverage and miss anything beyond the login page. We report the results of a preliminary experimental comparison with Monkey below.
Our operations logger performs network/cryptographic/file API instrumentations. It is resistant to obfuscation/packing for identifying Android SDK cryptographic APIs, supports  HTTP/S and non-HTTP protocols, supports (unobfuscated) 3rd-party encryption/decryption API, supports defragmentation of multi-part cryptographic functions and network packets. These features help us to understand a lot of custom-encrypted and non-HTTP traffic, and identify more privacy exposures compared to existing work.
Our data flow inspector detects privacy issues in the collected network traffic/files, by matching actual plaintext (collected by the app operations logger) and ciphertext from the network (after handling any IP defragmentation)---i.e., our reported exposures indeed happen during app runtime.
This helps us to avoid false positives. We reliably detect the use of weak cryptographic keys and algorithms; support various privacy-sensitive items (can be easily extended); support various encoding schemes, and nested encoding and encryption; support file detection within encrypted traffic; and distinguish between app and OS traffic. These features make the data flow inspector to accurately detect privacy and (potential) security problems.

\subhead{UI interactor vs.\ Android Monkey}
To compare the effectiveness of our UI interactor against Android Monkey (commonly used in past studies~\cite{50_ways,obfus_res,bakopoulou2020exposures}), we randomly chose 150 apps that exceeded the 5-minute threshold from our results. We set up two new experiments with a 10-minute threshold (following~\cite{50_ways}): in one experiment we used our UI interactor, and in another, we used the Monkey as the UI exerciser. We also configured Monkey to generate a large number of UI events, by setting a short interval of 0.3 seconds between events. Note that our interactor generates far fewer events---on average, 10 seconds per event, as we keep states to avoid duplicate events, perform text analysis, and use the online Google Translate service. We used the latest versions of the 150 randomly chosen apps, and 115 of them completed the analysis without any unexpected termination (in both Monkey and UI-interactor; we did not consider any partial results in this test).

In the end, our interactor spent about 7.4 minutes (444.36 seconds) on average for each app, while Monkey used the full 10-minute window (600 seconds) for each app. We compared our UI interactor and Monkey in terms of the detected various privacy-related items (a total of 24 types): on average, \ToolName\ detected approximately 6.5\% more apps with privacy leaks with our UI interactor compared to Monkey; see Fig.~\ref{fig:comp_pct_apps_data_types} (in the appendix). Most apps transmitted more privacy items when instrumented with our UI interactor. We also found that Monkey sent more duplicate items to the same host, or to new hosts (not detected by us). Most new hosts in Monkey received device names that appeared in the user-agent of an app's WebView pages that we intentionally avoided interaction with (e.g., ad windows, non-Google 3rd-party logins).

Additionally, we manually checked the support of login for these apps as the UI interactor can detect privacy leaks from app features available only after a successful login. We detected that 77/115 apps require authentication: 40 only supported app-specific authentication; 34 supported both Google and app-specific authentication; and 3 supported only Google sign-in. We succeeded to automatically login to 19 apps with Google sign-in and to 4 apps with app-specific registration/sign-in. %
This partial support of login helps us to explore more app features and related leaks compared to Monkey.

\subhead{Analysis time threshold: 5 vs.\ 10-minute window}
From the UI interactor vs.\ Monkey experiment, we estimate the undetected privacy leaks due to our 5-minute test window by comparing leaks that occur before and after the threshold. Overall, there are more leaks detected with a higher threshold, but the difference is not very significant. 6/115 apps sent the following privacy-related items after the 5-minute threshold: 1 sent the device name, 2 device email, 1 WiFi info (router BSSID, ESSID, and neighbor router ESSID), 1 dummy0 interface, 1 device name; i.e., 109/115 apps did not leak any new privacy-related items after the 5-minute threshold (and before the end of the 10-minute window). We also observed that most apps requiring over 5 minutes, are WebView apps with lots of pages/widgets. Note that, the analysis duration is configurable---a trade-off between coverage vs.\ total analysis time/resources.
\end{newtext}

\subhead{Limitations}
(1) Although we were able to identify PII information sent over the network with multiple forms of obfuscations (e.g., encryption, encoding, hashing), our results are a lower bound as we may not have identified traffic with more complex or unknown obfuscation techniques.
(2) Besides obfuscated PII, obfuscated method names may also reduce \ToolName's effectiveness, as we rely on method names for hooking possible encryption/decryption functions. Obfuscation tools such as ProGuard cannot modify method names in the Android cryptographic SDK (or any Android Framework APIs), allowing us to hook such functions successfully. However, these tools may hide from us the names of the custom-developed cryptographic functions, and as such, \ToolName\ cannot (automatically) find and hook these functions. From our measurement, we found a total of 119 apps that used non-SDK encryption; these apps either did not use obfuscation, or used some tools which did not obfuscate the method names. However, we could not measure how many apps with non-SDK encryption that we missed. Past studies measured the overall use of obfuscation tools by app developers, e.g., 24.92\% of 1.7M apps were obfuscated according to Wermke et al.~\cite{wermke2018large} (but no data on the use of non-SDK cryptographic implementations).
(3) Our instrumentations do not cover apps built solely using the native NDKs.
Instead, our methodology indirectly covers NDK functions that are wrapped in SDKs.
(4) The AndroidViewClient that we used to automate UI interactions, cannot handle animated UI elements (e.g., a button with an animation). We also do not handle UI elements created with third-party views (i.e., not extending \textit{View}~\cite{view} class) and images. Our support for authentication is also limited to Google sign-in and custom registrations, and our UI interactor currently does not complete steps that require verification via SMS/email for registration/login. For apps with unhandled UI elements and logins, we currently fail to detect privacy and security issues in features behind these elements or logins. (5)
As we do not know which apps, or which app features in an app may use non-standard/custom-encrypted channels, there is no guarantee that our UI interactor would trigger \emph{all} such covert channel related behaviors/features. We systematically go through all app UIs and trigger as many actions as we can to find these channels, if used by a target app.
(6) Our network instrumentation currently does not support HTTP/3 (QUIC), DTLS, and TLS without HTTPS.(7) Our Frida instrumentation works for the evaluated apps; however, if apps use advanced techniques, such as observing the memory map, our instrumentation can still be detected.

\section{Case Studies and Discussion} %
\label{sec:discussion}

In this section, we summarize privacy implications of the use of non-standard and covert channels to collect/send sensitive personal/device information. We also discuss the new security vulnerabilities introduced by these channels. We highlight such critical privacy and security implications using high-profile apps/SDKs as examples; see also Table~\ref{tab:nice_table} in the appendix. %

\subhead{Hiding privacy exposures} %
As we observed, a significant number of apps use non-standard and covert channels to hide the collection of PII/device information --- shared with their own app servers or third-party servers/trackers, or both. This may be due to increased scrutiny by the app markets, e.g., Google Play Protect~\cite{google_play_protect},
or due to the added privacy measures in newer versions of Android (10 and up); we cannot be certain about app developers' motivation on this.
However, such practices are certainly detrimental to user privacy.
We list a few examples below.
\\$\bullet$ Dailyhunt (\textit{com.eterno}, 100M+ installs), a top news app in India, sends users' contact list to its servers using an additional encrypted channel over HTTPS. It compresses and encrypts each contact's details using AES-128-CBC with a random key and a null IV, and sends the encrypted contacts to its server. The random key is also sent encrypted under a hard-coded RSA-1024 key. %
\\$\bullet$ SHAREit (\textit{com.lenovo.anyshare.gps}, 1B+ installs), an extremely popular app to securely share/manage large files, sends device GPS location to third-party adv/tracking domains (\textit{adcs.rqmob.com}~\cite{list_base_rules_block_apps}
and \textit{dt.beyla.site}~\cite{sitereview_dt_beyla_site})
under custom encryption over HTTP, and to \textit{mcds.wshareit.com} over regular HTTPS. For custom encryption, \textit{SHAREit} uses an AES-128 random key generated on the target device, which is sent encrypted via HTTP under a hard-coded RSA-1024 key.
Similarly, the Amazon Alexa (\textit{com.amazon.dee.app}, 50M+ installs) app sends the device email and the WiFi IP address (as cookie parameters) to their servers, only under custom encryption over HTTPS. %
This app is used to set up Alexa-enabled devices for automated tasks (e.g., creating shopping lists).
\\$\bullet$ With IPv6, the device interface's hardware MAC address is embedded in the IPv6 address~\cite{bssid-bh2021}, which is made available via the dummy interface (dummy0)~\cite{creating_dummy_interface}. Although the MAC address is randomized (by default from Android 10), the corresponding IPv6 address is fixed until the next reboot of the device~\cite{dummy_interface}, and as such, can be used to track users between device reboots (a relatively infrequent event).
This technique is apparently being used by \numc{202} apps, %
including very high-profile apps such as \textit{com.baidu.searchbox}, and \textit{com.paypal.android.p2pmobile}, where the apps explicitly collect the dummy0 interface information but use IPv4 for communication;
\printpercentoutof{14}{202} of them send such info via non-standard channels. %

\subhead{New security vulnerabilities}
We list example apps where new vulnerabilities resulted from the use of custom encrypted channels.
\\$\bullet$ UC Browser (\textit{com.UCMobile.intl}), a mobile browser with 500M+ installs, sends device information (e.g., device ID, operator, WiFi MAC, advertisement ID), and GPS location over a custom encryption channel under plain HTTP protocol with fixed-keys, and thus exposes the collected information to any network adversary.
\\$\bullet$ CamScanner (\textit{com.intsig.camscanner}), a widely-used app for document scanning (3.9M+ installations), encrypts a user's Firebase token using a random symmetric key, which is then encrypted using a hard-coded RSA-512 public key; the resulting ciphertext (the Firebase token and random key) is then sent to \textit{54.183.90.125:8090} and \textit{54.177.44.214:8090} using a non-standard protocol over TCP. We extracted the corresponding RSA-512 private key, and then we could recover the symmetric key and in turn, access the plaintext Firebase token,
just by collecting ciphertext from the network.
\\$\bullet$ We found \numc{22} apps that sent authentication tokens over a secure HTTPS channel initially, but then exposed  such tokens over an insecurely-implemented encrypted channel over HTTP or non-HTTP; e.g., \textit{com.peppermint.livechat.findbeauty} (a dating app, 5M+ installations) sent the user token over a non-HTTP channel that is AES-ECB encrypted with a hard-coded key.
\\$\bullet$ We also found 5 VPN clients use the shadowsocks~\cite{shadowsocks} protocol to receive free and premium server credentials from a proxy server through an encrypted channel. After decrypting the credentials on the device, it checks whether the user can authenticate and connect to premium or free servers. Since this check occurs at the client side, and \textit{\ToolName} can find the corresponding encryption parameters, we obtained connection information and credentials (e.g., server address, password) required to connect to the server.\\

\begin{newtext}
\subhead{To use or not to use non-standard channels and custom encryption} %
We checked several apps manually to understand their reasons  for using non-standard and custom encryption channels. The examples we observed do not clearly justify such channels, at least not in an obvious way (there may be deployment/operation constraints we are not aware of). Notable cases with raw TCP/UDP connections: Forex Event - Platform Trading (\textit{com.bonus.welfare}) uses a plain TCP channel to receive the latest shares and forex events; Modern Combat 5 (\textit{com.gameloft.android.ANMP.GloftM5HM}) uses a TCP channel apparently as a game control channel, and sends game server details and the user access tokens as plaintext; and Netspark Real-time filter (\textit{con.netspark.mobile}), a parental control app, sends real-time device activities (e.g., application-related events) to their server using a plain UDP channel.

The use of custom encryption should be avoided in general; as evident from our results, most app developers fail to use such encryption securely, e.g., about 87\% of apps used fixed keys for their symmetric cryptographic operations, where the ciphertext is indeed sent to the network. For specific app issues, we suggest the following fixes. For example, Recipes in Russian (\textit{ru.mediafort.povarenok}) could use their own public key to encrypt and store the device email on the shared storage; Comico (\textit{tw.comico}) could do the same to store their user authentication tokens; CamScanner's RSA-512, and both Dailyhunt and SHAREit's RSA-1024 keys could be replaced with a stronger one (e.g., RSA-2048); UC Browser could simply use HTTPS (instead of custom encryption over HTTP with a fixed key); for the 5 VPN apps that expose premium account checks at the end client side, these apps should perform the validation at their server-ends; and 22 apps that use custom encryption to share securely-established session tokens, should simply use HTTPS.

In the end, custom encryption is generally not the solution for any of the reasons that we observed---all of which can be easily met by Android's default crypto support. Besides using HTTPS properly for communication, app developers should rely on Android Keystore for local key management, and Android EncryptedFile and EncryptedSharedPreferences for securely storing local data~\cite{work_with_data_more_securely}.
To protect confidentiality of selected private content against third-party content-scanning/distribution services (e.g., allowing CDNs to scan HTTPS traffic), custom encryption may be used, but only under HTTPS (to limit any weakness of custom encryption to the CDNs, instead of any on-path attacker). To avoid the use of custom encryption over non-standard channels, e.g., AES-over-UDP/TCP, developers should instead choose QUIC.
\end{newtext}

\section{Related work}

\subhead{Privacy leakage via covert channels}
Side channels allow apps to access protected data/system resources, while with covert channels, an app can share permission protected data with another app that leaks permission-protected information. Reardon et al.~\cite{50_ways} automated the execution of \numc{88000} apps (at system and network levels), and monitored sensitive data sent over network traffic by apps, and scanned apps that should not have access to the transmitted data, due to lack of permissions.
The authors also reverse-engineered the respective components of apps to determine how the data was accessed, and studied the abuse from side and covert channels.
Examples from their findings include: 5 apps collect MAC addresses of connected WiFi base stations from ARP cache; an analytic provider (\textit{Unity}) obtained device MAC address using the \textit{ioctl} system call (42 apps were found to exploit this); third-party libraries from \textit{Baidu} and \textit{Salmonads}, wrote phone's IMEI to the SD card, and other apps that do not have permission to access the IMEI, can read from the SD card (13 such apps were found). They also found that \printpercentoutof{153}{88000} apps used hard-coded encryption keys.

Palfinger et al.~\cite{androtime} built a framework to identify timing side channels (e.g., via  querying installed apps, active accounts, files, browser logins) in Android APIs.
The leaked information can be used to fingerprint users, identify user habits or infer user identity.
Bakopoulou et al.~\cite{bakopoulou2020exposures}  intercepted the network traffic from 400 popular apps (with monkeyrunner), and performed manual/automated analysis to understand PII exposures. They found 29 apps exposed the ad ID and location info via UDP; 7 apps exposed Android ID, and another exposed username over plain TCP. %

We implement \ToolName\ to detect information leaks from non-standard and covert channels not reported in past studies --- e.g.,  malicious apps revealing neighbor's BSSID, obfuscation using nested encryption/decryption.
In addition to HTTP/HTTPS, we also capture traffic from other network protocols above TCP/UDP.
We found \printpercentoutof{2880}{2887} apps send/receive data over custom encrypted channels to/from the network; \printpercentoutof{414}{2880} of these apps used hard-coded keys on their communications. We also look for more fine-grained privacy-oriented sensitive information --- e.g., GPS coordinates with different accuracy, and user credentials.

\subhead{Obfuscation-resilient privacy leakage detection tools}
Mobile apps and ad libraries can leverage various obfuscation techniques (i.e., encoding, encryption, formatting) to hide the leakage of private information of users. Continella et al.~\cite{obfus_res} developed a tool (\textit{AGRIGENTO}) based on blackbox differential analysis (i.e., using a baseline, and observing the network traffic flow after modifying the sources of private information) for privacy leak detection resilient to underlying obfuscations. AGRIGENTO (implemented on Android 4) captures HTTP/HTTPS traffic using mitmproxy. The authors evaluated AGRIGENTO using the most popular 100 apps from Google Play, and identified 46 of them had privacy leaks; with manual inspection, the authors found that \printpercentoutof{4}{46} of those apps were false positives.
AGRIGENTO does not consider non-deterministic sources (e.g., one-time non-reusable keys, authentication tokens), and focuses on privacy leakages only from deterministic sources, i.e., Android ID, contacts, ICCID, IMEI, IMSI, location, MAC address, and phone number.
With AGRIGENTO, Continella et al.~\cite{obfus_res} also found false positives of specific sources of information leaked in a number of apps --- Android ID (5), IMEI (9) MAC address (11), IMSI (13), ICCID (13), location (11), phone number (16), contacts (13).
In contrast to AGRIGENTO, \textit{\ToolName} uses more comprehensive UI interactions, and relies on deep packet inspection; therefore, it can capture more privacy leaks from \emph{both} deterministic and non-deterministic sources.

\subhead{UI automation frameworks}
Past work~\cite{50_ways,obfus_res,syscall,explore_arm} has mostly relied on Appium~\cite{appium} and monkeyrunner~\cite{monkeyrunner} for Android UI automation. Appium uses app-specific scripts to drive automation relating to interactions with UI elements. %
Monkeyrunner solely uses random clicks on UI elements for automation. %
Dynodroid~\cite{dynodroid} focuses on processing automatic input. SmartDroid~\cite{smartdroid,dcdroid} automatically reveals UI-based trigger conditions of sensitive behaviors of Android apps, but it cannot interact with WebView (commonly used by recent apps). Patel et al.~\cite{patel2018effectiveness} found that random testing with monkeyrunner is extremely efficient, effective and leads to a higher coverage~\cite{patel2018effectiveness}. In contrast, Wang et al.~\cite{dcdroid} argue that monkeyrunner is unsuited for UI automation (for testing specific SSL/TLS vulnerabilities), as its random clicks do not precisely target the specific area on the UI.
They leverage AndroidViewClient for UI interactions (e.g., check a radio button, input content to a text box), based on the priority of a UI element; the priority depends on the  vulnerabilities in the SSL/TLS implementation. %
In contrast, we set the priority of UI element interaction in a list of clickable/fillable elements.
Our UI interactor is also built on top of AndroidViewClient, which has a better code coverage (e.g., accommodate interacting with UI elements that have non-English labels), not restricted to triggering UI elements associated to vulnerable SSL/TLS implementations, and
supports running on multiple devices.%

\subhead{Root detection evasion} %
We implement effective evasion mechanisms to bypass various root detection techniques incorporated by some apps. We use rule-based API hooking, and support both Android SDK and NDK based root detection. We surpass the capabilities built into common tools including \textit{RootCloak}~\cite{rootcloak}, \textit{RootCloak Plus}~\cite{rootcloakplus} and \textit{xCon}~\cite{xcon}, and handle more modern root detection measures; \textit{RootCloak} only supports up to Android \numc{6} and cannot bypass other tools/libraries like the \textit{RootBeer}~\cite{rootbeer}. We support Android 12 and can bypass more complicated techniques, including the latest version of \textit{RootBeer}, which is used
in \printpercentoutof{178}{6075} apps that trigger encryption/decryption APIs in our test.

\subhead{Defense against anti-reverse engineering techniques}
Android apps are prone to efficient reverse engineering, as apps are written in a high-level language (i.e., Java) that can be decompiled into simple bytecode~\cite{lim2016android}. To protect apps from reverse engineering,  past studies~\cite{ghosh2013shielding,xu2015toward,sun2015android} have %
discussed different obfuscation, dynamic code loading, packing, encryption and anti-debugging techniques and their detection and evasion.
We use Frida~\cite{Frida} API hooking to implement evasion techniques to protect against dynamic anti-reverse engineering bypassing detection techniques (e.g., the use of root access, package installer, mock location, certificate pinning, ADB).
Some of our techniques are more effective (e.g., mock location detection) compared to existing anti-evasion measures.

\begin{newtext}

\subhead{Summary of differences with existing work}
AGRIGENTO~\cite{obfus_res} is closest to our work; however, we cannot compare with it  experimentally (developed for now outdated Android 4). In terms of methodology, AGRIGENTO detects leakage of eight predefined, deterministic privacy-sensitive values: AndroidID, contacts, ICCID, IMEI, IMSI, location, MAC-address, and phone-number. %
We detect both fixed and dynamic values from deterministic/non-deterministic sources, as we have access to the plaintext corresponding to the full request. Also, due to the use of differential analysis in AGRIGENTO, there are a significant number of false positives.

Reardon et al.~\cite{50_ways} looked into unauthorized access and transmission of private data where an app does not have the necessary permissions. However, they did not address authorized/unauthorized privacy leakage via encrypted (beyond HTTPS) or non-HTTP channels, which is the focus of \ToolName. More concretely, from our results as summarized in Table~\ref{tab:content_types_on_channels}, everything except the ``Regular'' channels will be missed by other tools, except AGRIGENTO (albeit partial detection-only, as discussed above). Note that AGRIGENTO also does not consider security problems, and as such, issues reported under ``Credentials'' and ``Key Transmission'' in Table~\ref{tab:content_types_on_channels} will be missed. We also detect more privacy-sensitive data types (a total of 30 in the current implementation), compared to existing work (a total of 11 types in~\cite{obfus_res, 50_ways}). This can be attributed to our use of various known techniques in combination, such as bypassing runtime evasion, collecting non-HTTP traffic via tcpdump, logging non-SDK encryption/decryption methods and cryptographic APIs using rule based API-hooking (which can capture any runtime activities based on predefined criteria).

\end{newtext}

\section{Conclusion}
While considering the significant threat arising from non-standard and covert channels in the Android ecosystem, a better understanding of privacy exposures and security issues is necessary.  However, identifying privacy exposure via such channels is not straightforward.
Thus, users and app market providers would remain unaware of such privacy leakages, and security problems introduced by these channels.
We introduce \ToolName, a tool that can detect covert channels with multiple levels of obfuscation (e.g., encrypted data over HTTPS, encryption at nested levels). We also found security weaknesses caused by the use of custom-encrypted/covert channels (e.g., vulnerable keys and encryption algorithms).
With the findings and contributions from our study, we hope to spur further research in the context of non-standard and covert channels.

\begin{acks}
We are grateful to all anonymous CCS2022 reviewers for their insightful suggestions, comments, and guiding us in the final version of this paper. We also appreciate the help we received from the members of Concordia’s Madiba Security
Research Group. The third author is supported in part by an
NSERC Discovery Grant.
\end{acks}

\bibliographystyle{ACM-Reference-Format}
\bibliography{bibliography}

\begin{figure}[!htb]
\centering
  \includegraphics[scale=0.55]{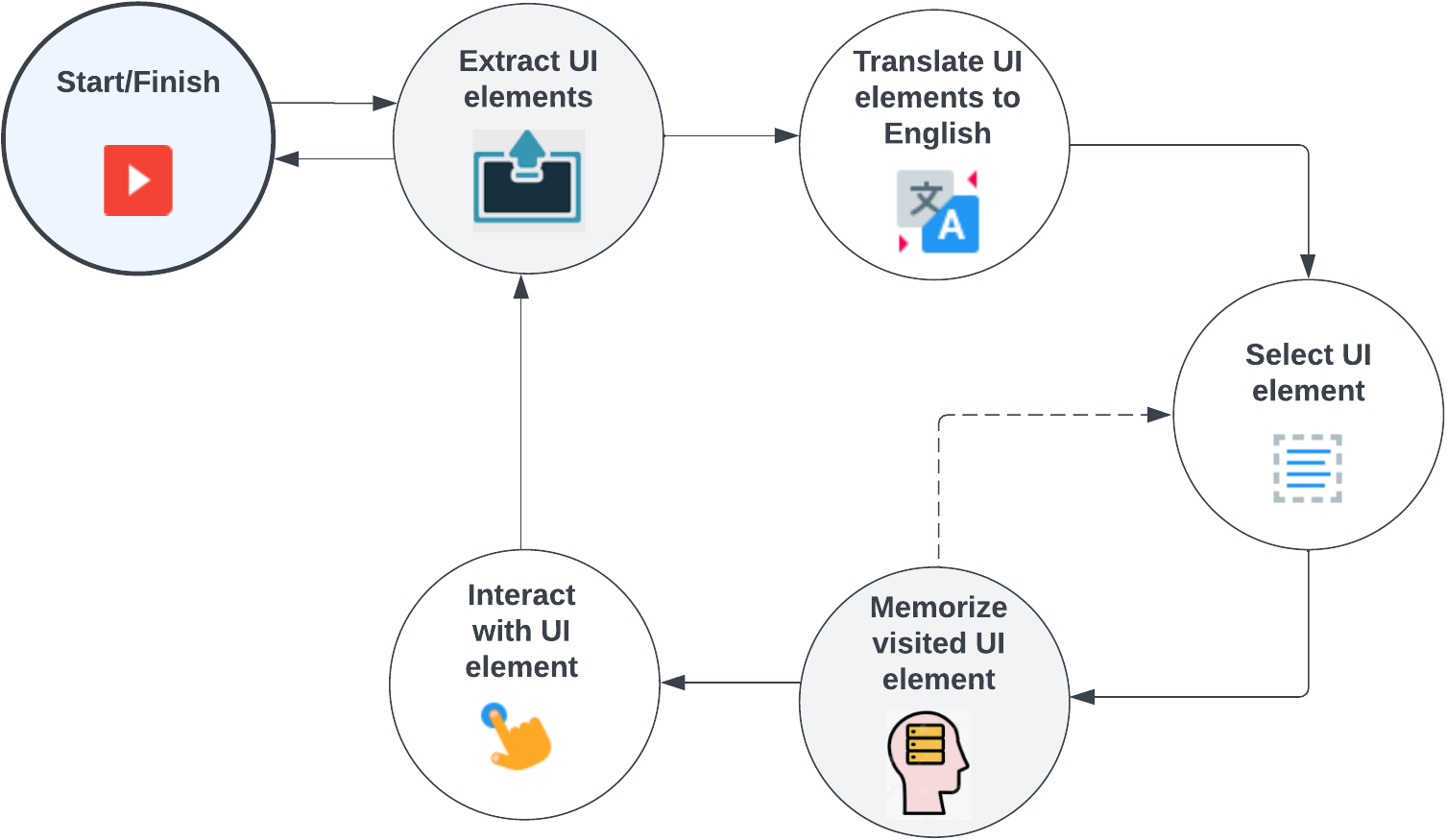}
  \caption{UI interactions using UI interactor}
   \label{fig:ui_interaction}
\end{figure} %

\begin{table}[!htb]
\begin{tabular}{  p{2.5cm}|p{3.25cm}  }
\textbf{Input field} & \textbf{Input value}  \\
\hline
username & admin88888888 \\
email & mymail@email.com \\
e-mail & mymail@email.com \\
name & admin88888888 \\
mobile number & 9158888888 \\
phone & +888888888888 \\
password & P4ss@88888888 \\
code & 9188 \\
pin & 1370 \\
city & Montreal \\
id & 9188888888 \\
How old & 45 \\
YYYY & 1991 \\
search & search88888888 \\
\hline
\end{tabular}
    \caption{Input list used for user interactions --- priority of the input fields decrease from the first row to the last row}

    \label{tab:input_list}
\end{table}

\begin{table}[!htb]
\resizebox{\linewidth}{!}{
\begin{tabular}{ l r r l l}
\textbf{Remote domain} & \textbf{\# Apps} & \textbf{\# Installs} & \textbf{Company} & \textbf{Country} \\
\hline
appsflyer.com&1386&28.40b+&AppsFlyer&USA\\
supersonicads.com&284&7.90b+&ironSource&Israel\\
scorecardresearch.com&193&2.70b+&Comscore&USA\\
isnssdk.com&147&4.00b+&ByteDance&China\\ %
uc.cn&100&1.40b+&Alibaba Group&China\\
pangle.io&72&1.90b+&ByteDance&China\\ %
batch.com&68&0.20b+&Batch&France\\
mopub.com&63&2.00b+&AppLovin&USA\\
qq.com&35&0.58b+&Tencent&China\\ %
umeng.com&32&0.47b+&Umeng&China\\
\hline
\end{tabular}}
    \caption{Top-10 domains that receive traffic from apps with an additional layer of encryption (the cumulative app installation number is given in billions).}
    \label{tab:encrypted_destinations}
\end{table} %

\begin{table}[!htb]
\begin{tabular}{ p{4cm}|p{1.9cm}|p{1.4cm} }
\textbf{App name} & \textbf{Amplification Ratio} & \textbf{\#installs} \\
\hline
com.popa.video.status.download& \numc{19314} & 5M+\\
com.peach.live& \numc{19314} & 10M+\\
com.download.funny.online& \numc{19314} & 10M+ \\
com.asiainno.uplive.aiglamour& \numc{12098} & 50M+\\
com.meiqijiacheng.sango& \numc{12021} & 10M+\\
com.facechat.live& \numc{11622} & 5M+\\
com.kitty.android& \numc{6328} & 10M+\\
mozat.rings.loops& \numc{6328} & 5M+\\
com.yiyo.android& \numc{5287} & 10M+\\
com.peachpro.live& \numc{5286} & 5M+\\
com.bela.live& \numc{4362} & 10M+\\
\hline
\end{tabular}
    \caption{Apps with top-10 amplification ratios} %
    \label{tab:amplification ratio}
\end{table}

\begin{table}[!htb]
\begin{tabular}{ l|l }
\textbf{Java API}	& \textbf{Native API} \\
\hline
Socket::connect&connect\\
DatagramSocket::connect&recv\\
DatagramSocket::send&send\\
DatagramSocket::receive&sendto\\
SocketChannel::connect&recvfrom\\
SocketChannel::open&accept\\
DatagramChannel::connect&listen\\
SmartFox::connect&bind\\
\hline
\end{tabular}
    \caption{Java and native network APIs}
    \label{tab:instrumentation_network_api}
\end{table}

\appendix
\clearpage
\section{Anti-reverse engineering bypass} \label{sec:runtime_anti_rev}
We use Frida to implement the following components to protect our app execution against common runtime evasion techniques.

\subhead{Root detection}
We explored several commonly used root detection bypass frameworks (\textit{RootCloak}~\cite{rootcloak}, \textit{UnRootBeer}~\cite{unrootbeer}, \textit{Fridantiroot}~\cite{fridantiroot} and \textit{Objection}~\cite{objection}), and found that those are still detected by some apps. %
Thus, to cover more apps, we build a more generic anti-root detection module, supporting both SDK/NDK based detection (which can also bypass the latest version of \textit{RootBeer}~\cite{rootbeer}, among others).
We use a simple rule-based API hooking technique: all methods with ``rooted'' in their names that return a boolean value, are modified to always return \texttt{false}.

\begin{sloppypar}
\subhead{Package installer detection}
Apps
can check their package installer app to ensure that they are installed from Google Play. Apps that are installed using other package installers (i.e., marketplaces) may behave unexpectedly (e.g., close app). This is problematic for our analysis when an app is installed from a local APK file. We thus hook the \textit{getInstallerPackageName} function (from  \textit{android.app.ApplicationPackageManager} and \textit{android.content.pm.InstallSourceInfo} classes), and set the package installer name to always return \textit{com.android.vending}, which is the installer name for Google Play.
\end{sloppypar}

\subhead{Mock location detection}
We develop an app to mock GPS coordinates
to fixed values. Since apps can identify the use of mock GPS feature, we hook \textit{isMock} and \textit{isFromMockProvider} methods (from \textit{android.location.Location}
class), to always return \texttt{false} that hides our GPS mock service.

\begin{sloppypar}
\subhead{Frida detection}
We use Frida for API hooking, which runs a TCP server on port 27042. We observe that some apps (e.g., \textit{com.scotiabank.banking}, a Canadian bank) attempt to connect to this port, to check whether Frida is running, and if so, terminate the app execution.
To solve this problem, we hook the \textit{connect} function of the \textit{Bionic}~\cite{bionic} library instead to connect to port 1, making the default port of Frida-server to appear as closed to the target app.
By hooking the low-level \textit{connect} function of the Bionic library instead of the \textit{connect} function of \textit{java.net.Socket} class, we cover all network sockets (i.e., achieve better API coverage). %
\end{sloppypar}

\subhead{Certificate/public key pinning}
We use a third-party script~\cite{Frida_android_unpin} to bypass certificate/public key pinning, which works on most common TLS libraries, e.g., OkHttp3, TrustManager, SSLSocketFactory,
OpenSSL, and Tustkit.

\subhead{ADB detection}
Apps may check for ADB status and behave differently, affecting our analysis as we use ADB to communicate between our test desktop and target devices.
We thus hook all \textit{getter} method in the \textit{android.provider.Settings.Global} class to change the value of \textit{adb\_enabled} ADB status to 0 (i.e., to make ADB to appear as disabled).

\begin{sloppypar}
\subhead{Lock detection}
For automation, we disable the passcode in our target devices. However, some apps (e.g., \textit{ca.bc.gov.id.servicescard}, a Canadian government app) change their behavior if the device does not use a passcode. Therefore, we hook the \textit{isDeviceSecure} method in the \textit{anandroid.app.KeygaurdManager} class, which checks the device passcode, and we always return \texttt{true} to avoid detection.
\end{sloppypar}

\begin{table}[htb]
\begin{tabular}{ l|r | r|r}
\textbf{Channel/media}	&  \textbf{Symmetric} &  \textbf{Public key} &  \textbf{Non-SDK} \\
\hline
HTTPS	&2492 &	424 & 101\\
HTTP &	255 &	206	&27\\
Non-HTTP &	23 &	8&	0\\
File &	24 &	1 &	1 \\
\rowcolor{gray!25}
Overall& 2597& 598& 119\\
\hline
\end{tabular}
    \caption{Types of encryption used by apps in network traffic/files } %
    \label{tab:prevalance_enc}
\end{table}

\begin{table}[htb]
    \centering
    {\small{
\begin{tabular}{ l| r r| r r }
 \multirow{2}{*}[0ex]{\textbf{Content type}} &\multicolumn{2}{c|}{\textbf{Fixed keys (HTTPS)}}&\multicolumn{2}{c}{\textbf{Fixed keys (HTTP)}} \\
&{\cellcolor{gray!25}Multi-Dev}&{\cellcolor{gray!25}Single-Dev}&{\cellcolor{gray!25}Multi-Dev}&{\cellcolor{gray!25}Single-Dev} \\
\hline
Device& 1328& 1824& 51& 66\\
Network& 1156& 1473& 9& 16\\
Network Location& 13& 13& 2& 3\\
GPS Location& 29& 36& 7& 7\\
User Assets& 36& 43& 3& 5\\
Credentials& 171& 226& 11& 14\\
Key Transmission& 8& 12& 6& 5\\
\hline
\end{tabular}
}
}
    \caption{Content types sent over HTTPS/HTTP from apps using fixed symmetric encryption/decryption keys (Multi-Dev = across multiple devices, Sing-Dev = single device)}
    \label{tab:onsecure_key_mgmt}

\end{table}

\begin{table}[htb]
    \centering
    {
\begin{tabular}{l r r r r}
\textbf{Algorithm} & \textbf{HTTP} & \textbf{HTTPS} & \textbf{Non-HTTP} & \textbf{Storage} \\
\hline
RSA\-768& 3& 0 & 0 & 0 \\
RSA\-512& 1& 58 & 1 & 0 \\
RSA\-384& 0& 1 & 0 & 0 \\
DES& 25& 83 & 0 & 7\\
3DES& 15& 22 & 0 & 0 \\
RC4& 0& 3& 0& 0 \\
ECB& 216& 606 & 13 & 1 \\
\hline
\end{tabular}
}
    \caption{Weak ciphers used by apps in storage media and network communications
    }
    \label{tab:weak_cipher_protocols}
\end{table}

\section{Other network security threats}
\label{sec:netsec-misc}
We found \printpercentoutof{2183}{12598} and \printpercent{128}{12598} apps sent sensitive information over HTTP and non-HTTP channels, respectively.
We also found that \printpercentoutof{86}{12598} apps used the UDP protocol for non-standard use (i.e., DNS, NTP, QUIC are excluded).
The server of the apps that use UDP protocol can be vulnerable to UDP-based amplification attacks.
To measure such vulnerability, we collect all UDP packets that are generated by the app from the tcpdump (pcap) data, but skip packets destined for a remote server on ports \numc{53}, \numc{123}, and \numc{443} to avoid getting standard protocol packets (i.e., QUIC, NTP, DNS). Then, we extract their payload and destination address; afterwards, we resend their payload in new UDP sockets to their destination. %
Since UDP does not guarantee delivery of packets, we wait for \numc{5} seconds after resending the UDP packets, to ensure a response from the server.
Then, we compare our sent and received payloads to determine if the received payload is larger than the sent payload (i.e., by evaluating the amplification ratio). We perform our measurement two weeks after our analysis, to determine whether these attacks tend to continue for days and are independent of the sessions created during our interactions in the initial experiment. In the end, we found that \printpercentoutof{10}{86} of the app servers had an amplification ratio over 4300, which could be exploited for DDoS attacks (see Table~\ref{tab:amplification ratio}).

\clearpage

\begin{figure*}[!htb]
\centering
  \includegraphics[scale=0.7]{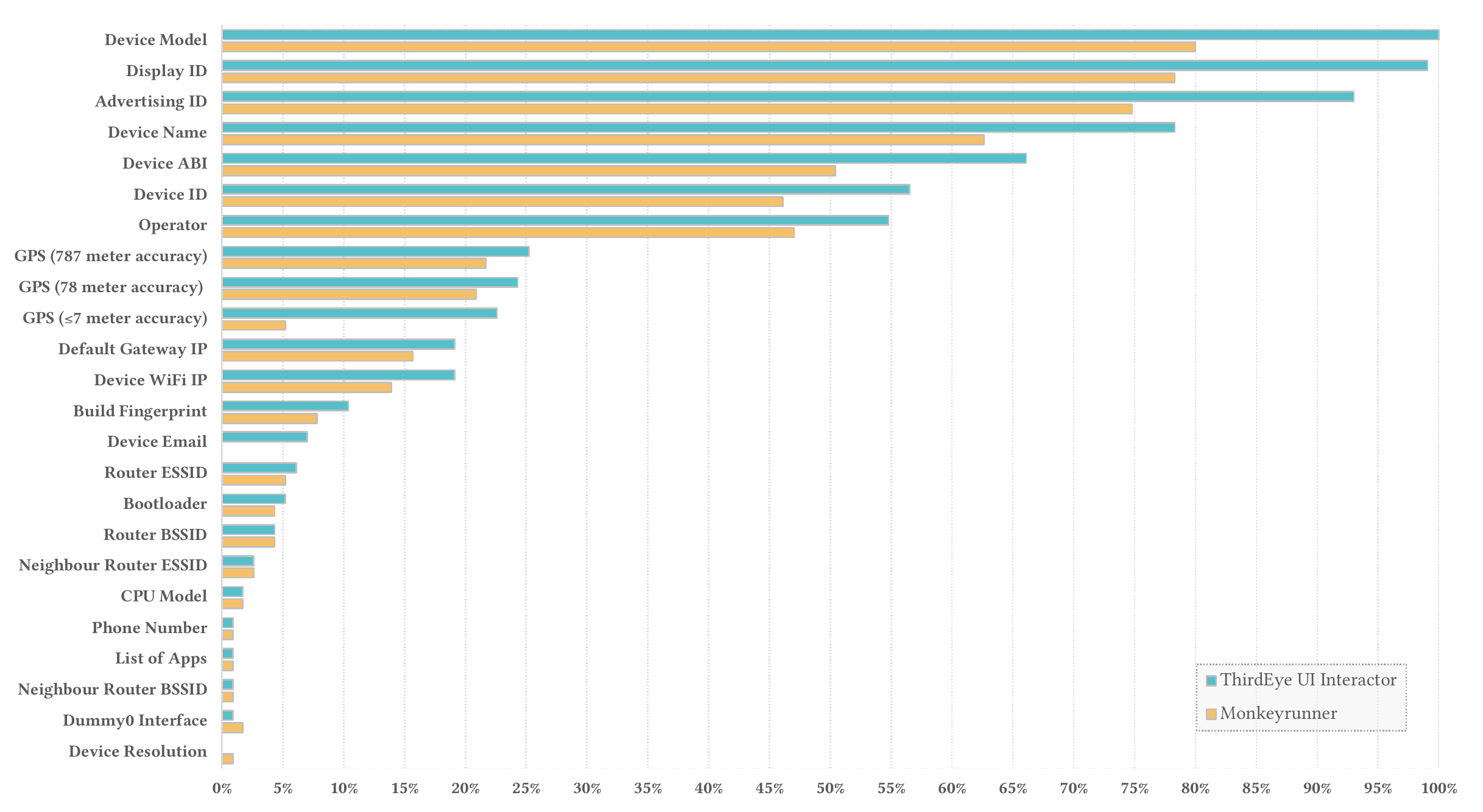}
  \caption{Percentage of apps exposing various data types between \ToolName\ and Monkey (out of \numc{115} apps)}
   \label{fig:comp_pct_apps_data_types}
\end{figure*}

\begin{figure*}[!htb]
\centering
  \includegraphics[scale=0.715]{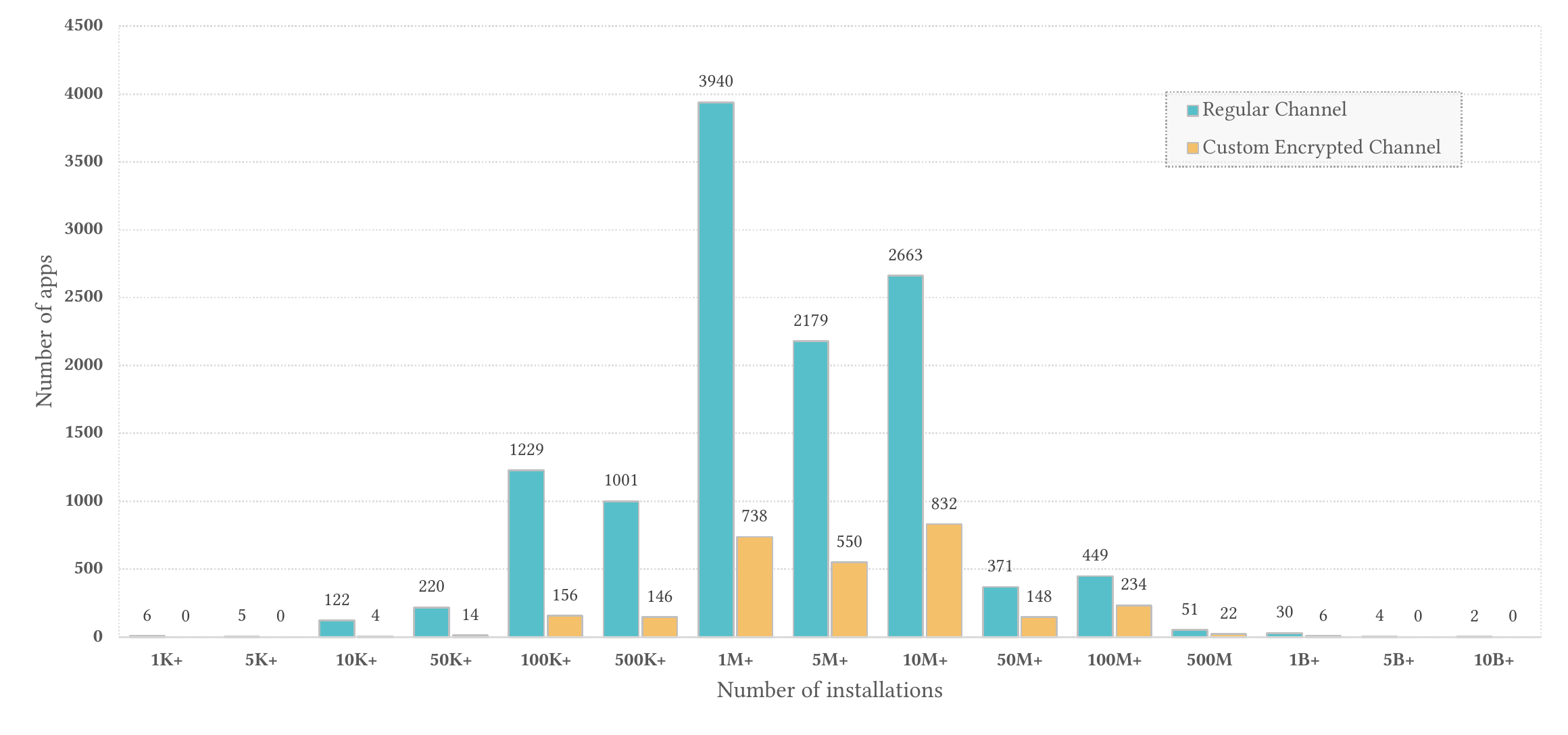}
  \caption{Distribution of our apps in terms of installation numbers (as of Jan.\ 30, 2022) and their use of regular vs.\ custom encryption channels. We collected the dataset between Nov.\ 25, 2021 and Jan.\ 6, 2022, and 326 apps were removed from Google Play as of Jan.\ 30, 2022 (which we omit here).} %
   \label{fig:comp_pct_apps_installs}
\end{figure*}

\begin{figure*}[!htb]
\centering
  \includegraphics[scale=0.75]{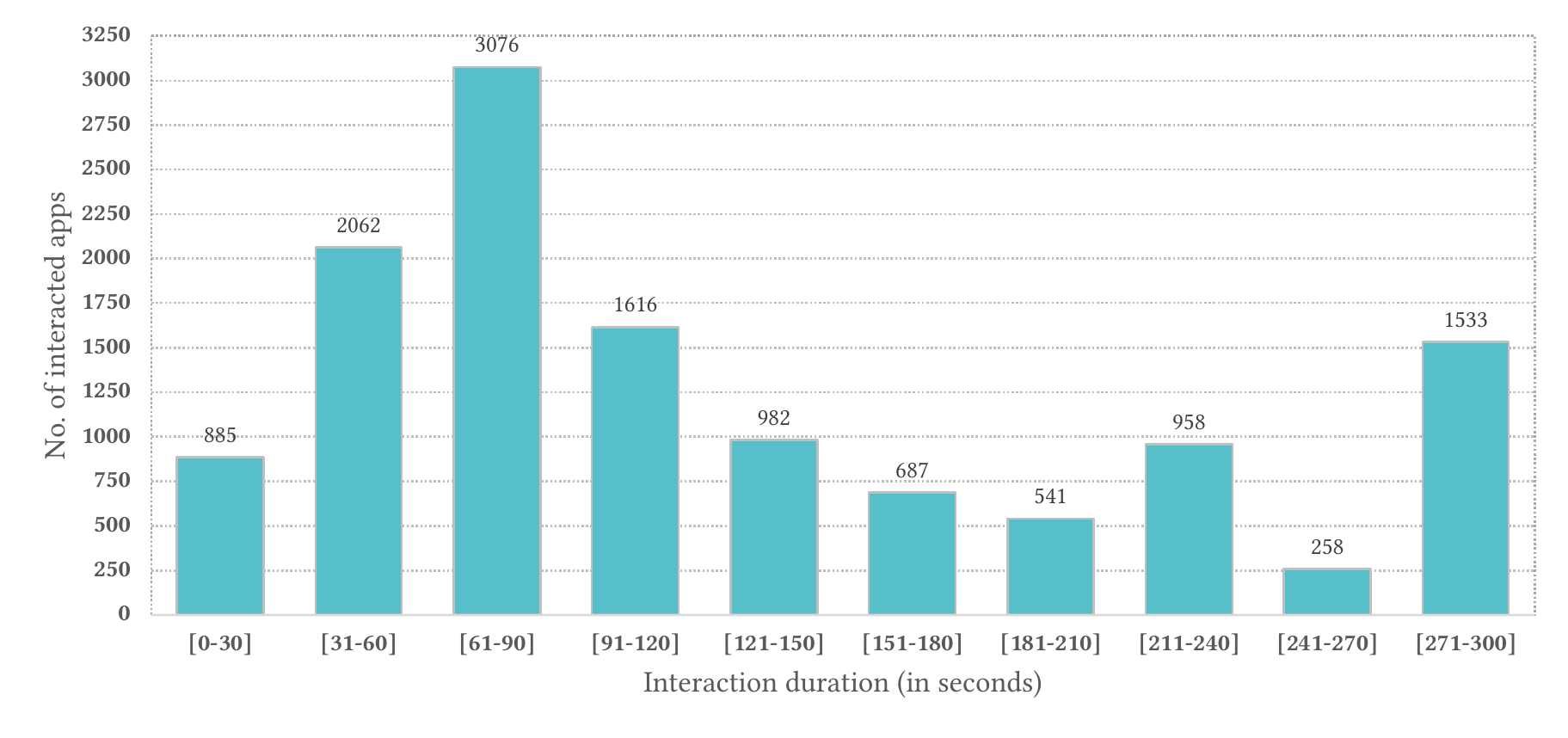}
  \caption{Number of apps that completed their UI interactions within different time intervals (up to 300 seconds)}
   \label{fig:app_interaction_timings}
\end{figure*}

\begin{table*}[!htb]
    \centering
    \begin{tabular}[t] { p{2.5cm}|p{3.25cm} }
        \textbf{Keyword} & \textbf{Exclude keyword(s)}  \\
        \hline
none of the above & - \\
not now & - \\
get code & - \\
approv & - \\
<phone number>* & - \\
read & already \\
scan & - \\
resend & - \\
woman & - \\
women & - \\
female & - \\
global & - \\
<username>* & - \\
country & - \\
<country> & - \\
<state> & - \\
18 & - \\
skip & - \\
gmail.com & - \\
start & startup \\
agree & disagree, agreement \\
next & - \\
allow & disallow \\
continue & continue with \\
confirm & - \\
accept & not accept \\
checkbox & - \\
without & - \\
google & share \\
register & - \\
sign & help, can \\
login & help, can, facebook, google \\
\hline
\end{tabular}%
 \hspace{15pt}%
\begin{tabular}[t] { p{2.5cm}|p{3.25cm} }
        \textbf{Keyword} & \textbf{Exclude keyword(s)}  \\
        \hline
log in & help, can, facebook, google \\
sign in & help, can, facebook, google \\
enable & - \\
submit & - \\
retry & - \\
again & - \\
current & - \\
i don & - \\
find & - \\
guest & - \\
ok & facebook, book \\
test & - \\
done & - \\
use & user \\
go & forgot, good, google \\
an account & - \\
later & - \\
comment & - \\
quick & - \\
close & - \\
send & - \\
english & - \\
dismiss & - \\
cancel & - \\
great & - \\
navigation & - \\
settings & - \\
free & - \\
limit & - \\
tap & - \\
click & - \\
icon & - \\
\hline
\end{tabular}%
    \caption{Keyword list used for user interactions --- the priority of keywords decreases from the first row to the last row, and from the left-side table to the right-side table; $*$ the values for the keywords are obtained from Table~\ref{tab:input_list}.}
    \label{tab:keyword_list}

\end{table*}

\begin{table*}[!htb]
\rowcolors{2}{gray!5}{gray!30}
\centering
\begin{tabular}{ >{\raggedright}p{5cm}|p{8cm}| c }
\rowcolor{white}
\textbf{Application name (package)} & \textbf{Implication} & \textbf{Installs} \\
\hline
SHAREit (com.lenovo.anyshare.gps) & Sends device GPS coordinates to third-party domains \textit{adcs.rqmob.com} and \textit{dt.beyla.site} by using a custom encryption channel over HTTP %
&1,000M+ \\
UC Browser-Safe (com.UCMobile.intl) &Sends device info and GPS coordinates over an insecure HTTP custom encrypted channel&500M+ \\
Picsart Photo \& Video Editor (com.picsart.studio) & Sends the GPS coordinates to tracking domains \textit{analytics.picsart.com}, \textit{ads.mopub.com}, \textit{conversions.appsflyer.com}; the last domain used a custom encryption channel & 500M+ \\
Likee (video.like) & Sends network device proxy, default gateway, and WiFi interface addresses  only over a custom encryption channel over HTTPS & 500M+ \\
CAIXA (br.com.gabba.Caixa) & Sends device GPS coordinates and neighbor router information to a third-party party domain \textit{evg.dnofd.com} (this domain appeared in 6 Brazilian governmental developers in our result) & 100M+ \\
Dailyhunt (com.eterno) & Sends device contact only over a custom encrypted channel over HTTPS & 100M+ \\
Phoenix Browser (com.transsion.phoenix) & Sends user's search and browsing history and on-device info over an insecure custom encrypted channel over HTTP & 100M+ \\
JOOX Music (com.tencent.ibg.joox) & Sends on-device information including own routers to \textit{101.33.47.68:8081} over a custom encryption channel over TCP & 100M+\\
Phone Master (com.transsion.phonemaster) & Sends device GPS coordinates and neighbor router information to a third-party domain \textit{evg.dnofd.com} (this domain appeared in 9 developers in our result) & 100M+ \\
Amazon Alexa (com.amazon.dee.app) & Sends device email and WiFi IP address only by using custom non-SDK encryption channel over HTTPS as cookie parameters & 50M+ \\
Garena Li\^{e}n Qu\^{a}n Mobile (com.garena.game.kgvn) & Sends on-device information including own router info to \textit{101.33.47.68:8081} over a custom encryption channel over TCP; its encryption methods are also packed & 50M+ \\
Baidu (com.baidu.searchbox) &Sends various on-device information including neighbor and own router info and GPS coordinates only via a custom encryption channel over HTTPS & 5M+ \\
MOMS (com.moms.momsdiary) & Sends its \emph{pushtoken} (used for notification) to the server encrypted (using a non-SDK algorithm) under a fixed key over HTTP. An attacker can replace a victim's push-token with their own to receive the victim's notifications & 0.5M+ \\

\hline
\end{tabular}
    \caption{Examples of sensitive information sent by Android apps over non-standard/covert channels;  note that for Baidu, we provide the number of installs from Google Play, but Baidu is primarily installed in China through Chinese app markets.}
    \label{tab:nice_table}
\end{table*}

\end{document}